\documentclass[aps,prd,preprint,groupedaddress, showpacs, amsfonts,amssymb,amsmath, floatfix]{revtex4}

\usepackage{bm}
\usepackage{graphicx}

\begin{document}

\title{Coupling of Linearized Gravity to Nonrelativistic Test Particles:
Dynamics in the General Laboratory Frame}

\author{A.~D.~Speliotopoulos}

\email{adspelio@uclink.berkeley.edu}

\affiliation{Department of Physics, University of California,
Berkeley, CA 94720-7300}

\author{Raymond Y. Chiao}

\email{chiao@physics.berkeley.edu}

\homepage{physics.berkeley.edu/research/chiao/}
{}
\affiliation{Department of Physics, University of California,
Berkeley, CA 94720-7300}

\date{March 8, 2004}

\begin{abstract}

The coupling of gravity to matter is explored in the linearized gravity
limit. The usual derivation of gravity-matter couplings within the
quantum-field-theoretic framework is reviewed. A number of
inconsistencies between this derivation of the couplings, and the
known results of tidal effects on test particles according to
classical general relativity are pointed out. As a step towards
resolving these inconsistencies, a General Laboratory Frame fixed on the
worldline of an observer is constructed. In this frame, the
dynamics of nonrelativistic test particles in the linearized gravity
limit is studied, and their Hamiltonian dynamics is derived. It is
shown that for stationary metrics this Hamiltonian reduces to the
usual Hamiltonian for nonrelativistic particles undergoing geodesic
motion. For nonstationary metrics with long-wavelength gravitational
waves (GWs) present, it reduces to the Hamiltonian for a nonrelativistic
particle undergoing geodesic \textit{deviation}
motion. Arbitrary-wavelength GWs couple to the test particle through a
vector-potential-like field $N_a$, the net result of the tidal
forces that the GW induces in the system, namely, a local velocity
field on the system induced by tidal effects as seen by an observer in
the general laboratory frame. Effective electric and magnetic fields,
which are related to the electric and magnetic parts of the Weyl
tensor, are constructed from $N_a$ that obey equations of the same
form as Maxwell's equations . A gedankin gravitational
Aharonov-Bohm-type experiment using $N_a$ to measure the  interference
of quantum test particles is presented.

\end{abstract}


\maketitle


\section{Introduction}

At first glance, it would seem that the Hamiltonian dynamics of
nonrelativistic, classical test particles in the linearized gravity
limit has been thoroughly studied, and is well understood. Indeed, in
this limit gravitational waves (GWs) are often treated as simply a spin-2
gauge field propagating in flat Minkowski spacetime \cite{Feyn}, and the
coupling of GWs to matter would seem to follow naturally. This
determination would be premature; we show in this paper that such an
approach obfuscates the underlying physics of the system, and
overlooks the surprising links between gravitational waves, vector
potentials, and gauge symmetries.

Much of our current understanding of the coupling of matter to gravity
comes from attempts at constructing quantum gravity (QG) \cite{Feyn,
deWa, deWb, deWc}, and from the theory of quantum fields in curved
spacetime (QFCS) \cite{BD, Wald_0}. Once the Lagrangians for various
elementary particles\textemdash both gauge and nongauge\textemdash
were determined in flat Minkowski spacetime, their extension
to curved spacetimes was a natural next step. To make this extension
of the flat spacetime Lagrangians to curved spacetimes, a number of
seemingly natural assumptions were typically made then, and are still
being made now. The expectation is that the experience and intuition
gained from constructing quantum field theories (QFTs) in flat
spacetime will serve as useful guides in constructing QFTs in curved
spacetimes. Thus, flat spacetime Lagrangians for bosonic fields are
promoted to curved spacetimes by replacing the Minkowski metric
with the metric for a curved spacetime, the partial derivative with
the covariant derivative, and the Lorentz-invariant integration
measure with the general-coordinate-transformation-invariant integration
measure. The extension of fermionic fields, such as spin-1/2 and -3/2
fermions, follow in much the same way once a tetrad frame is
chosen. The Hilbert action from classical general relativity (GR) is
used for the gravity component of the theory, and the metric for the
spacetime $g_{\mu\nu}$ is identified with the gravitational field's
degree of freedom. A classical background metric $g^0_{\mu\nu}$ for
the spacetime is chosen\textemdash in \cite{Feyn} it was flat
Minkowski spacetime, and in \cite{deWa, deWb, deWc} it was either an
asymptotically flat spacetime or a spacetime with finite spatial
extent\textemdash and the propagating component of the gravitational
field is extracted from the theory by considering fluctuations
$h_{\mu\nu}$ about $g^0_{\mu\nu}$ combined with a suitable gauge
(coordinate) choice. These fluctuations\textemdash representing
gravitational waves (GWs) in GR and gravitons in QG\textemdash are then
expanded about $g^0_{\mu\nu}$ assuming that $h_{\mu\nu} = g_{\mu\nu} -
g^0_{\mu\nu}$ are small compared to $g^0_{\mu\nu}$, and are
subsequently treated as simply another spin-2 non-Abelian gauge field
propagating in the background spacetime. By also expanding the metric
terms about $g^0_{\mu\nu}$ in the Lagrangians for the matter fields,
one obtains terms that couple GWs\textemdash or gravitons in
QG\textemdash with matter. These interaction terms would then seem to
be fixed by the field's corresponding flat spacetime Lagrangians
combined with the standard prescription for promoting them to curved
spacetimes.

As natural, and as straightforward, as the above prescription is for
determining the coupling of matter to gravity, it nevertheless makes a
number of implicit assumptions. When one tries to reconcile these
assumptions with classical GR, a number of troubling inconsistencies
become immediately apparent.

The first implicit assumption is that the measuring apparatus does not
play a role in the theory. That is, when calculating the various effects
caused by the interaction of gravity with matter\textemdash such
as, say, the scattering cross section of GWs\textemdash one
does not have to explicitly include the measuring apparatus. This
assumption is certainly true for all the other forces of nature; the
existence of opposite or canceling ``charges''  for the EM, weak, and
strong forces ensures that one can, in principle, screen out these
forces, thereby making the measuring apparatus unaffected by
them. This  is \textit{not} true for gravity, however; one cannot
screen out the gravitational force. Why, then, should one \textit{not}
include the measuring apparatus explicitly in the construction of QG or of
QFCS? Indeed, we shall show here that one \textit{must} include the
effects of gravitation on the apparatus in order to obtain physically
correct results.

One may argue that the scattering processes considered in quantum
gravity occurs at very short length scales\textemdash the Planck
length\textemdash and the presence of any measuring apparatus will
have a negligibly small effect. However, we would expect from the
correspondence principle that classical gravity results could be
obtained\textemdash in some limit\textemdash from the quantum
theory; indeed, the construction of QG \cite{Feyn, deWa, deWb, deWc}
makes explicit use of the classical theory. And in classical GR it is
well known that the inclusion of the measuring apparatus\textemdash
along with the observer\textemdash is crucial to understanding the
dynamics of certain time-varying general relativistic systems
involving tidal forces.

Consider, for example, an isolated observer and a classical test particle
initially at rest some distance away from him.  While both the
observer and the particle do not move spatially with respect to one
another, they are both physical objects that move along their
respective geodesics. If a GW now passes through the system, \textit{tidal}
forces will of course shift the position and the velocity of the
particle. However, these same tidal forces will \textit{also} shift the
position and the velocity of the observer; the observer cannot be
isolated from the effective tidal forces caused by the
GW. Thus, the observer cannot measure the motion of the particle
independently of his own motion; he can only measure the
\textit{relative} motion of the particle with respect to himself. For
GWs in the long-wavelength limit, the particle appears to the
observer to undergo geodesic \textit{deviation} motion [Eq.~(35.12) of
\cite{MTW}], and not geodesic motion as one might first
expect. Indeed, a simple derivation of the geodesic deviation equation
in this limit is to take the geodesic equation for the observer,
subtract from it the geodesic equation of the particle, and expand
the result in the distance separating the two geodesics.

Based on this simple example we would expect that any physically
measurable response of matter to the scattering of GWs calculated by
using either QG or QFCS should include the effect of the GW on the
observer. By extension, we would expect that in order to be consistent
with classical GR, the construction of QG or QFCS should explicitly
include the observer and his measurement apparatus from the very
beginning.

It may also be argued that, as with the other forces, explicit
inclusion of an observer would be formally correct, but not required;
the lack of its inclusion would not materially affect any
calculation. This argument would also be in conflict with
general relativity, however.

Consider once again the simple system described above. When the
geodesic deviation equations of motion are solved, one finds that the
observed \textit{tidal} response of the test particle to the passage of the
long-wavelength GW is proportional to the distance separating the
observer from the test particle; as long as the long-wavelength
approximation holds, the further away the particle is from the observer,
the larger its response to the GW. This ubiquitous response of classical
matter to the passage of a GW is exploited in various GW detectors
such as the Weber bar and LIGO (\textit{L}aser \textit{I}nterferometry
\textit{G}ravitational-wave \textit{O}bservatory); the larger the
detector, the larger its response to the passage of a GW. The
characteristic size $L$ of the detector \textit{does} play a role in
the response of the system to the GW.

Suppose that either QG or QFCS is used to calculate the response of a
Weber bar or LIGO to passage of a GW through the system. It would be
natural to use the complex scalar field $\Psi$ to describe the
system. Using the standard approach outlined above, it is
straightforward to see that to lowest order, the coupling between a GW
propagating in Minkowski spacetime with $\Psi$ is $\sim h_{\mu\nu}
\partial^\mu \Psi^\dagger \partial^\nu \Psi$. Even in the
long-wavelength limit, this interaction term does not explicitly
depend on the size of the system. Moreover, it is difficult to
see how such size dependence can be generated by this term.

The second implicit assumption made in \cite{Feyn} and \cite{deWa,
deWb, deWc} is that one can always find a global time axis\textemdash
and thereby construct a global coordinate system\textemdash in the
curved spacetime. This is certainly possible for flat Minkowski
spacetime, which is often used as the background spacetime. It is also
possible for the asymptotically flat manifold that DeWitt considers in
\cite{deWa}. However, we know from classical GR that it is not
possible to find a coordinate system with a \textit{global} time axis
in general.

The Minkowski spacetime and the asymptotically flat
spacetimes\textemdash along with the various black hole
spacetimes\textemdash are \textit{stationary} spacetimes. In these spacetimes
one can always choose a frame where the metric does not depend
explicitly on the time coordinate. Consequently, one can always
construct a \textit{global} timelike Killing vector, which can be used
by all observers in the spacetime as their time axis (except, perhaps,
at the event horizon or at an essential singularity). This Killing
vector can then be used to construct what DeWitt termed the ``preferred
frame'', or as Hawking and Ellis termed it, a ``special frame''
\cite{hawking} that all observers in that spacetime can agree to use.

Timelike Killing vectors\textemdash and global frames\textemdash do
not exist in general, however. Importantly, they do not exist in the
presence of a GW. Instead, each observer must choose his own local
proper time axis, and construct his own local proper coordinate system
from it. Consequently, one can only
measure the \textit{relative} motion between observers. This is the
underlying physical reason why, in the example given above, one ends up
with the geodesic \textit{deviation} equation of motion \cite{MTW}
in the long-wavelength limit for GWs, instead of the geodesic equation
of motion.

Although the above point is made very elegantly in the beginning of
Chap.~4 of \cite{hawking} for \textit{classical} GR, it is relevant
on the quantum level as well. As pointed out in both Chap.~3 of
\cite{BD} and Chap.~3 of \cite{Wald_0}, the construction of Fock
spaces for quantum fields in a curved spacetime is frame dependent;
different choices of coordinates result in unitarily
\textit{inequivalent} Hilbert spaces. Thus, only for such spacetimes
as the stationary and De Sitter spacetimes (considered also by DeWitt
in \cite{deWa}), where there is a ``preferred frame'', will it be possible
for all observers to agree on what constitutes a particle state. It
does not exist in general (see \cite{Wald_0} for a discussion of the
relevance of the concept of ``particles'' in general spacetimes).

As dissimilar as the two above implicit assumptions may be appear to
be on the surface, they are nonetheless intimately connected. The experimental
measurement of \textit{any} physical quantity requires an
\textit{operational} choice of origin, and a \textit{local} orthogonal
(tetrad) coordinate system. As any measurement is done through a
physical apparatus, this mathematical choice of coordinate systems is
fixed on a real physical object. The inclusion of the observer in the
theory is thus equivalent to a choice of local coordinates; a choice
of local coordinates must be equivalent to the inclusion of the
physical observer.

Although we have pointed out in the above a number of inconsistencies
between results from classical GR on the one hand, and QG and QFCS on
the other, the goal of this paper is \textit{not} to present a
reformulation of either QG or QFCS; we leave that task to
future research. We instead address the issues raised by the two
assumptions in the above by focusing on the dynamics of a much
simpler system: the nonrelativistic, classical test particle in weak
gravity. As simple as this system may be, especially when compared
to the counterexamples we have listed above, many of the issues that
we have raised above appear here as well. Fundamentally, what is at
issue here is the appropriate choice of coordinates; this is an
inherent aspect of classical GR, and is not due to a subtlety in the
quantum theory.

An analysis based on the dynamics of classical test particles has the
added advantage of having limiting cases that have either been
experimentally verified, or are in the process
of being verified. In one limit, the E\"otv\"os experiment, the
advancement of the perihelion of Mercury, the deflection of light by
the Sun, and the gravitational redshift are all calculable within the
usual dynamics of test particles in stationary spacetimes based on the
geodesic equation. In the other limit, the response of Weber bars and
LIGO to the passage of GWs is calculable within the dynamics of test
particles based on the geodesic \textit{deviation} equation
\cite{MTW}. The result of the analysis in this paper \textit{must}
agree with these two limits; this serves as a stringent test of the
validity of the approach we have taken and the coordinate system we
have constructed.

In the literature most analyses of the dynamics of test particles in
curved spacetime are done in the same vein as the construction of QG
and QFT in curved spacetimes, and are a direct generalization of the
usual techniques for deriving Hamiltonians from Lagrangians for
particles in flat spacetimes. One starts with the usual
\textit{geodesic} action for the test particle moving in an arbitrary
curved spacetime with a given metric $g_{\mu\nu}$. Time
reparametization invariance of the action is broken either by choosing
an explicit time coordinate, or by introducing a mass-shell constraint
(by hand or through a Lagrange multiplier). Choosing $x^\mu$ as the
general coordinate, the canonical momentum $p_\mu$ is calculated
from the Lagrangian. The Hamiltonian $H_{SF}$ ($SF$ for ``standard
formalism'') is then constructed from this $p_\mu$ and the Lagrangian
in the usual way. An analysis similar to this was followed by DeWitt
in his 1957 paper \cite{deW1}, albeit in much more detail, and in 1966
he applied the nonrelativistic limit of $H_{SF}$ for charged test
particles to the analysis of the behavior of superconductors in the
Earth's Lense-Thirring field \cite{deW2}.

It would seem that all we would have to do is to take the
nonrelativistic limit of $H_{SF}$. However, the form of $H_{SF}$ is
dramatically different from the Hamiltonian for test particle motion
derived in \cite{ADS1995} based on the geodesic \textit{deviation}
equations of motion.  As with the case of QG and QFCS the same
troubling questions come to the fore at this point: Where is the
observer? What are physical quantities such as the position $\vec x$
and velocity $\vec v$ of the particle measured with respect to? What
frame has been implicitly chosen by this analysis? Is this frame
physical? We know from the observer\textendash test-particle example
given above that these are not fatuous questions. Rather, they
directly address the underlying physics.

It is certainly true that in some specific cases\textemdash such as the
presence of a weak GW in the system\textemdash one can treat the
time-varying part of the metric as a perturbation of one of the known
stationary metrics; this time-varying piece would then be reflected as
a perturbation on the Killing vector. One could then use the usual
coordinate system for these spacetimes\textemdash augmented by the
inclusion of the observer and his coordinate system\textemdash and
calculate $H_{SF}$ in the usual way. Doing so will not
elucidate the underlying physics, however, and it is difficult to see
how the geodesic \textit{deviation} equations of motion arises in this
approach.

The approach we shall take instead in studying the dynamics of
nonrelativistic test particles in the linearized gravity limit will
be to construct a general coordinate system that builds in the
essential physics from the very beginning. Since \textit{relative}
measurements between the observer and
the particle always make physical sense, they are used as the
foundation of our construction; the special case of stationary metrics
will naturally be included. Specifically, we follow the considerations
of \cite{Synge} and \cite{deF} (see also \cite{Thorne} in connection
with the coordinate system used in the analysis of LIGO): Every
physical particle travels along a worldline $\mathit{\Gamma}_c(\tau)$
with tangent vector $c^\mu$ (which does not need to be a geodesic) in
the spacetime manifold $\mathbb{M}$. Every measurement of the physical
properties of the test particle by an observer must be done using an
experimental apparatus. The observer\textemdash along with his
apparatus\textemdash must propagate along his own worldline
$\mathit{\Gamma}_u(\tau)$ with tangent vector $u^\mu$. Consequently,
every physical measurement of the particle is done \textit{relative}
to the motion of an observer. In particular, in measuring the position
of the particle, one measures the distance \textit{separating}
$\mathit{\Gamma}_c(\tau)$ and $\mathit{\Gamma}_u(\tau)$; in measuring
the $4$-velocity of the particle one measures of the \textit{relative}
velocity of the particle with respect to the observer \cite{MTW}.

Implementation of the above considerations proceeds quite
naturally. As the observer prepares to take measurements on the test
particle, he first chooses a local orthonormal coordinate system. In
curved spacetimes, this involves the construction of a local tetrad
frame \cite{Synge}. Naturally, this coordinate system will be fixed,
say, to the center of mass of his experimental apparatus, and will
thus propagate in time along the worldline $\mathit{\Gamma}_u(\tau)$
as well. The observer uses the coordinate time of the physical
apparatus to measure time, which, because he will not be moving
relative to the apparatus, is also his proper time. Thus the time axis
of the coordinate system he has chosen will always lie tangentially to
$\mathit{\Gamma}_u(\tau)$. The position\textemdash which can be of
finite extent\textemdash of the test particle is measured with respect
to an origin fixed on the apparatus, and is the shortest distance
between this origin and the particle. However, because the apparatus
travels along its worldline, the origin of the coordinate system will also
travel along a worldline in $\mathbb{M}$. Later, when the rate of
change of the position of the particle is measured at two successive
times, the \textit{relative} $4$-velocity of the particle with respect
to the apparatus will naturally be obtained. Thus, the
observer constructs his usual laboratory frame that extends across his
experimental apparatus, but now incorporating the nontrivial
local curvature of $\mathbb{M}$. We call this frame the general
laboratory frame (GLF).

Local coordinate systems fixed to an observer have been
constructed before. The Fermi normal coordinates (FNC) were
constructed in the 1920's by Fermi \cite{Fermi}, and the Fermi Walker
coordinates (FWC) were constructed in \cite{Synge}. While an observer
can use either set of coordinates, both make assumptions and
approximations that drastically limit their usefulness. The
FNC\textemdash a direct implementation of the
equivalence principle\textemdash are constructed so that the
Levi-Civita connection $\Gamma^\alpha_{\mu\nu}$ vanishes identically
along the worldline of the observer; only when one moves off the
worldline does the curvature dependent terms begin to appear
\cite{ManMis}. For the FWC, the restrictions on
$\Gamma^\alpha_{\mu\nu}$ are somewhat relaxed, but certain components
of $\Gamma^\alpha_{\mu\nu}$\textemdash such as
$\Gamma^{\hat{a}}_{\hat{0}\hat{b}}$ where $\hat{a}$, $\hat{b}$ are
spatial indices\textemdash still vanish along the worldline. Once
again, when one moves off the worldline curvature terms appear in the
form of the Riemann tensor and its derivatives. In both, one effectively
makes a derivative expansion in the Riemann curvature tensor
\cite{Mashhoon1, Mashhoon2, Ni, FNC5}.

In both FNC and FWC systems, choices for the value of
$\Gamma^\alpha_{\mu\nu}$\textemdash a gauge choice\textemdash have
been made, and in both systems such gauge choices are inconsistent with the
usual transverse-traceless (TT) gauge for GWs. While it is possible to
study the interaction of GWs with test particles in these coordinate
systems (see \cite{FNC1, FNC2, FNC3} for FNC and \cite{Mashhoon3}
for FWC), doing so is cumbersome. For example, it has only recently been
established that the TT gauge for GWs is compatible with the FNC
\cite{FNC4}, but only in the long wavelength limit; the two are
incompatible when the wavelength becomes smaller than the size of the
experimental apparatus. In our construction of the GLF, no such
restrictions on $\Gamma^\alpha_{\mu\nu}$ are made within the
linearized gravity approximation. Thus, when we consider the case of
GWs interacting with nonrelativistic particles, the TT
gauge\textemdash or any other gauge\textemdash can be directly
taken. Moreover, we do not make any restrictions on how rapidly the
Riemann curvature tensor varies, and therefore are not restricted to
only the long-wavelength limit. This enables us to study the effects of
arbitrary-wavelength GWs on the motion of nonrelativistic test
particles in large systems.

It is here in our study of test particle dynamics that we obtain our
most surprising result:  Even though the underlying GW is a spin-$2$
tensor field, in the weak gravity, slow velocity limit, the GW acts
on the particle through a local velocity field $N_a$. This velocity
field\textemdash which is an integral of the Ricci rotation
coefficients\textemdash couples to the test particle as though it was
a vector potential for a spin-$1$ \textit{vector} field (see also
\cite{Chiao_0} for the additional terms that the Ricci rotation
coefficients introduce in fermionic condensed matter systems and their
implications), and its origin is the \textit{tidal} nature of the
forces that the GW induces on the test
particle. It has the same properties as a vector potential: Like the
vector potential for the EM field $A_a$, $N_a$ is a transverse field
satisfying the wave equation. It is a frame-dependent field with the
local Galilean group as its gauge group. Effective ``electric'' and
``magnetic'' fields can be constructed from $N_a$ in the usual way,
and they are solutions to a set of partial differential equations that
have the same form as the Maxwell equations since they are directly
related to the electric and magnetic parts of the Weyl tensor, and
thus to components of the Riemann curvature tensor. The equations of
motion for the nonrelativistic particle have the form of a Lorentz
force with the mass of the particle playing the role of the charge. As
required, these equations reduce to the usual geodesic deviation
equations \cite{MTW} in the long-wavelength limit.

The rest of this paper is organized as follows. In Sec.~II we
construct explicitly the GLF and its coordinates using a tetrad frame
fixed to the worldline of the observer. The velocity of a test
particle in the GLF is derived in the nonrelativistic
limit. In Sec.~III we use these velocities to construct the
action, and then the  Hamiltonian for the test particle in the GLF. We
show that for stationary $\mathbb{M}$ this Hamiltonian reduces to
DeWitt's Hamiltonian, and for long-wavelength TT GWs propagating in a
flat background it reduces to the Hamiltonian \cite{ADS1995} derived from
the geodesic equations of motion. In Sec.~IV, we study the properties
of the velocity field $N_a$ introduced in Sec.~3 for arbitrary GWs, and
construct effective electric and magnetic fields from it. These fields
are shown to obey equations that have the same form as Maxwell's
equations, and they are used to derive the equations of motion for a
test particle. An Aharonov-Bohm-type interference effect for quantum
test particles that is shown to follow from the effective vector
potential $N_a$ can be found in Sec.~V along with other concluding
remarks. In Appendix A we present a brief  review of the tetrad
and linearized gravity formalisms, while in Appendix B, we derive the
nonintegrable phase factor $\exp\{i(m/\hbar)\oint N_A dX^A\}$ for the
gravitational Aharonov-Bohm-type interference effect.

\section{Construction of the GLF}

As usual, $g_{\mu\nu}(x)$ is the
metric on the curved spacetime manifold $\mathbb{M}$ with a signature
$(-1,1,1,1)$.  Greek indices run from $0$ to $3$, and they denote the
coordinates $x^\mu$ for a general coordinate system on
$\mathbb{M}$. We will, however, be working primarily in one specific
tetrad frame, and we will use capital Roman letters running from $0$ to $3$
for the spacetime indices in this frame. (A summary of well-known
results for linearized gravity and tetrad frames is given in
Appendix A.) We reserve lowercase Roman letters running from $1$
to $3$ for spatial indices in the tetrad frame, and careted lowercase
Roman letters for spatial indices in the general coordinate frame. A
worldline with a timelike tangent vector $u^\mu$ parameterized by
$\tau$ is denoted as $\mathit{\Gamma}_u(\tau)$; this worldline needs
not be a geodesic. Spacelike geodesics, with tangent vectors $\chi^\mu$
and parameterized by its arclength $s$, are denoted by
$\mathit{\gamma}_\chi(s)$, and null geodesics with
tangent vectors $\pi^\mu$ parameterized by its arclength $\sigma$
are denoted by $\mathit{\gamma}_\pi(\sigma)$.

The construction of the GLF for the observer\textemdash being a
   specific choice of general tetrad frames that is fixed onto the
   worldline of the observer\textemdash is fairly standard. It must,
   however, be done without knowing the specific form of the
   underlying metric of $\mathbb{M}$. Indeed, the local metric at any
   given time is \textit{determined} by making local measurements. We
   are aided in this construction by three observations. First and
   foremost, we note that the observer does not need a coordinate
   system that is nonsingular over \textit{all} of $\mathbb{M}$; such
   a coordinate system is  known not to exist in general. All that is
   needed is a coordinate system that is nonsingular within the region
   of $\mathbb{M}$ where the observer makes experimental
   measurements. Second, we are working in the linearized gravity
   limit. This assures that we do not have to concern ourselves with
   coordinate singularities, and we can take curvature effects as
   perturbations on the flat spacetime metric. Finally, we are
   primarily interested in the effect of linearized gravity on
   \textit{nonrelativistic} test particles; in this limit,
   incorporation of causality effects in the construction of the GLF
   simplifies dramatically.

Let us consider an observer $\mathcal O$ with worldline
   $\mathit{\Gamma}_u(\tau)$. To perform experimental measurements at
   some time $\tau_0$, $\mathcal O$ constructs a local orthogonal
   coordinate system centered on his experimental apparatus by choosing
   a tetrad frame $\{{}_oe_A^{\>\>\>\>\mu}(\tau_0)\}$, a set of
   orthogonal unit vectors such that $\eta_{AB} =
   {}_oe_A^{\>\>\>\>\mu}(\tau_0)\>\>
   g_{\mu\nu}\vert_{\mathit{\Gamma}_u(\tau_0)}\>\>
   {}_oe_B^{\>\>\>\>\nu}(\tau_0)$, where $\eta_{AB}$ is the usual
   Minkowski metric and $g_{\mu\nu}\vert_{\mathit{\Gamma}_u(\tau_0)} =
   {}_oe_{A\mu}(\tau_0)\>\> \eta^{AB}\>\>
   {}_oe_{B\nu}(\tau_0)$ is the metric for $\mathbb{M}$ at
   $\mathit{\Gamma}_u(\tau_0)$. We use the presubscript $o$ (for
   ``observer'') on ${}_oe_0^{\>\>\>\>\mu}$ to emphasize that at this
   point the frame  only exists on $\mathit{\Gamma}_u(\tau)$. Unlike the
   general tetrad frame, we require that $u^\mu(\tau_0) =
   {}_oe_0^{\>\>\>\>\mu}(\tau_0)$; the time axis of the frame at
   $\tau_0$ lines up with the worldline of the  observer. As usual, tetrad
   indices are raised and  lowered by $\eta_{AB}$, and general
   coordinate indices are raised and lowered by $g_{\mu\nu}$.

For the coordinate system at subsequent times we have to transport
${}_oe_A^{\>\>\>\>\mu}$ along $\mathit{\Gamma}_u(\tau)$ in such a way
that that ${}_oe_0^{\>\>\>\>\mu}$ always points along $u^\mu$. If
$\mathit{\Gamma}_u(\tau)$ were a geodesic, we would only need to
parallel transport $e_A^{\>\>\>\>\mu}$ along it. However, because we
are interested in nongeodesic worldlines we must instead use
Fermi-Walker transport, a generalization of parallel transport that
subtracts the nongeodesic motion of $\mathit{\Gamma}_u(\tau)$
from the transport of ${}_oe_A^{\>\>\>\>\mu}$. For any vector $v^\mu$
and a tangent vector $\chi^\mu$ to some worldline $\mathit{\Gamma}_\chi$,
the Fermi-Walker transport of $v^\mu$ along $\mathit{\Gamma}_\chi$ is
\begin{equation}
\frac{D^F v^\mu}{\partial \tau} = \frac{D v^\mu}{\partial \tau} -
\left(v_\nu \frac{D \chi^\nu}{\partial \tau}\right) \chi^\mu +\left(v_\nu
\chi^\nu\right) \frac{D \chi^\mu}{\partial \tau},
\label{1}
\end{equation}
where as usual parallel transport along $\mathit{\Gamma}_\chi$ is
\begin{equation}
\frac{Dv^\mu}{\partial \tau}= \frac{\partial v^\mu}{\partial \tau} +
\Gamma^\mu_{\alpha\beta} \chi^\alpha v^\beta,
\label{2}
\end{equation}
and $\Gamma^{\mu}_{\alpha\beta}$ is the connection on
$\mathbb{M}$.

By the Fermi-Walker transport of ${}_oe_A^{\>\>\>\>\mu}(\tau)$ along
$\mathit{\Gamma}_u(\tau)$, we find at each time $\tau>\tau_0$,
\begin{equation}
\frac{D^F {}_oe_0^{\>\>\>\>\mu}(\tau)}{\partial \tau}
\equiv 0.
\label{3}
\end{equation}
Not surprisingly ${}_oe_0^{\>\>\>\>\mu}(\tau)$ automatically undergoes
Fermi-Walker transport. The spatial tetrads, on the other hand, do not,
and are solutions of the linear partial differential equations
\begin{equation}
0 = \frac{D {}_oe_a^{\>\>\>\>\mu}(\tau)}{\partial \tau} -
\left({}_oe_a^{\>\>\>\>\nu}(\tau)\frac{D{}_oe_{0\nu}(\tau)}{\partial
\tau}\right) {}_oe_0^{\>\>\>\>\mu}(\tau),
\label{4}
\end{equation}
with the appropriate initial condition at $\tau = \tau_0$.

\begin{figure}
\begin{center}
\includegraphics[width=0.5\textwidth]{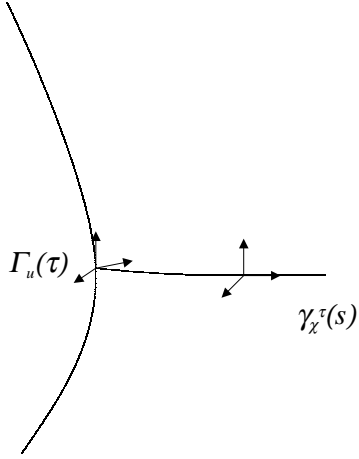}
\end{center}
\caption{\label{Fig-2}Parallel transport of $e_A^{\>\>\>\>\mu}$
off $\mathit{\Gamma}_u(\tau)$ and along a spacelike geodesic
$\mathit{\gamma}_\chi^\tau(s)$.}
\end{figure}

To extend $\{{}_oe_A^{\>\>\>\>\mu}(\tau)\}$ off $\mathit{\Gamma}_u(\tau)$
we once again use the Fermi-Walker transport of $e_A^{\>\>\>\>\mu}$, but now in
directions perpendicular to $\mathit{\Gamma}_u(\tau)$. At any fixed
time $\tau$, let $\mathit{\gamma}_\chi^\tau(s)$ be a spacelike
geodesic such that $u^\mu\chi_\mu = 0$ and $\mathit{\gamma}_\chi^\tau(0) =
\mathit{\Gamma}_u(\tau)$ (see Fig.~\ref{Fig-2}).
We include the superscript $\tau$ on $\mathit{\gamma}_\chi^\tau(s)$ to
denote the implicit dependence of $\mathit{\gamma}_\chi^\tau(s)$ on
$\tau$. For geodesics, Fermi-Walker transport is equivalent to
parallel transport, and $e_A^{\>\>\>\>\mu}(\tau,s)$ are solutions of
\begin{equation}
0=\frac{D e_A^{\>\>\>\>\mu}(\tau,s)}{\partial s},
\label{5}
\end{equation}
along $\chi^\mu$ with the initial condition $e_A^{\>\>\>\>\mu}(\tau,0)
= {}_oe_A^{\>\>\>\>\mu}(\tau)$ for each time $\tau$. It is
straightforward to show from Eq.~$(\ref{5})$ that
$e_A^{\>\>\>\>\mu}(\tau,s)\>e_{B\mu}(\tau,s) =
\eta_{AB}$. Consequently, we can now consider $e_A^\mu$ to be a vector
field on $\mathbb{M}$ such that
\begin{equation}
g_{\mu\nu} = e^A_{\>\>\>\>\mu}\> e_{A\nu},\qquad
\eta_{AB}=e_A^{\>\>\>\>\mu}\> e_{B\mu}.
\label{6}
\end{equation}

While the above defines a set of orthonormal vectors for the observer,
we still have to construct explicit coordinates for this frame. As
mentioned in the Introduction, observer-based coordinates have been
constructed before \cite{Fermi}, \cite{Synge}, although they were not
explicitly constructed in a tetrad frame. Although the FNC and FWC
are constructed using a series of approximations that make them
unsuitable for our purposes, a number of the fundamental concepts used
in their construction nevertheless carry over to our construction. In
particular, Synge \cite{Synge} introduces the notion of the world
function to construct both coordinate systems (see also
\cite{deF}). This \textit{extended} object is a scalar function that
measures the length squared between two points connected by a geodesic
on $\mathbb{M}$ that are separated by a \textit{finite} distance. It
serves as a two-point correlation function that measures the net effect of
the differences in local curvature between the two points. Moreover,
because the world function is a length\textemdash and thus a scalar
invariant\textemdash it is expandable in terms of the Riemann
curvature tensor and its derivatives, thereby avoiding many
coordinate-dependent artifacts.

Useful though the world function may be, with the tetrad frame
constructed above we have a more direct method of
constructing coordinates. (This method is similar to the approach
followed in \cite{FNC5} for the FNC.) Like Synge, our method makes
use of an extended object between two points on $\mathbb{M}$, and we
explicitly introduce a test particle $\mathcal P$ with worldline
$\mathit{\Gamma}_c(\tau)$ that is close enough to
$\mathit{\Gamma}_u(\tau)$ for its physical properties to be measured
by $\mathcal O$'s experimental apparatus. We ask what the coordinates
of this particle are in $\mathcal O$'s frame. To be consistent with
$\mathcal O$'s experimental measurements, we parameterize
$\mathit{\Gamma}_c(\tau)$ by the proper time of $\mathcal O$, not
$\mathcal P$. Then let $X^A(\tau)$ be the position of $\mathcal P$ at
any time $\tau$ in the observer's frame. $X^A(x): x^\mu \to X^A$ can
also be considered as a coordinate transformation from the general
coordinates $x^\mu$ to the tetrad frame at any time $\tau$, which in a
small neighborhood $\mathcal{U}_\mathcal{O}$ of $\mathcal{O}$ is
\begin{equation}
\eta_{AB}\> dX^A dX^B = \frac{\partial X^A}{\partial x^\mu}
\bigg\vert_{\mathcal{U}_{\mathcal{O}}} \frac{\partial X^B}{\partial
x^\nu}\bigg\vert_{\mathcal{U}_{\mathcal{O}}} dx^\mu dx^\mu,
\label{7}
\end{equation}
so that from Eq.~$(\ref{6})$,
\begin{equation}
\frac{\partial X^A}{\partial x^\mu}\vert_{\mathcal{U}_{\mathcal{O}}} =
e^A_{\>\>\>\>\mu},\qquad
\hbox{or}\qquad \frac{\partial x^\mu}{\partial
  X^A}\vert_{\mathcal{U}_{\mathcal{O}}} =e_A^{\>\>\>\>\mu}.
\label{8}
\end{equation}
Taking the derivative of the first equation in Eq.~$(\ref{8})$ with
respect to $x^\nu$, we find that the integrability condition:
$\partial_\nu e^A_\mu = \partial_\mu e^A_\nu$. This
condition only holds within $\mathcal{U}_\mathcal{O}$ (see
Eq.~$(\ref{64})$). To extend $X^A$ off $\mathcal{O}$'s worldline, we
note the following.

\begin{figure}
\begin{center}
\includegraphics[width=0.5\textwidth]{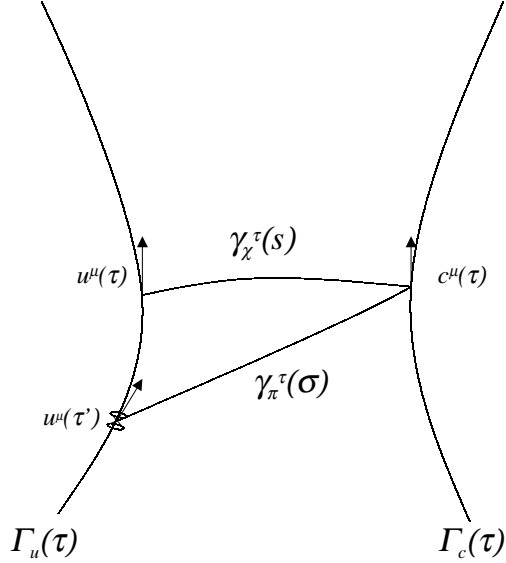}
\end{center}
\caption{\label{Fig-3}A sketch of the worldlines
$\mathit{\Gamma}_u(\tau)$ and $\mathit{\Gamma}_c(\tau)$ of observer
$\mathcal O$ and test particle $\mathcal P$, respectively, with the
spacelike $\mathit{\gamma}_\chi^\tau(s)$ and null
$\mathit{\gamma}_\pi(\sigma)$ geodesics used to construct $X^A$
shown. The end points of both geodesics are fixed onto specific points
on $\mathit{\Gamma}_u(\tau)$ and $\mathit{\Gamma}_c(\tau)$. The
simultaneity with respect to the observer $\mathcal O$ in time along
$\mathit{\gamma}_\chi^\tau(s)$ is an approximation that only holds in
the linearized gravity and nonrelativistic limits.}
\end{figure}

Solutions of Eqs.~$(\ref{8})$ are clearly path
dependent. For spatial components $X_a$, we consider again the
spacelike geodesic $\mathit{\gamma}_\chi^\tau(s)$, but now connecting
$\mathit{\Gamma}_u(\tau)$ to $\mathit{\Gamma}_c(\tau)$ such that
$\mathit{\gamma}_\chi^\tau(0) = \mathit{\Gamma}_u(\tau)$ and
$\mathit{\gamma}_\chi^\tau(s_1) = \mathit{\Gamma}_c(\tau)$. (See
Fig.~\ref{Fig-3}.)
By integrating Eq.~$(\ref{8})$ along this geodesic,
we obtain the spatial coordinates of the test particle
\begin{equation}
X_a(\tau) =
\int_{\mathit{\gamma}_\chi^\tau(0)}^{\mathit{\gamma}_\chi^\tau(s_1)}
e_{a\mu}(\tau,s)\chi^\mu(\tau,s) ds \equiv
\int_{\mathit{\gamma}_\chi^\tau(s)} \bm{e}_a,
\label{9}
\end{equation}
as a straightforward extension of the tetrad framework to
coordinates. We have made use of differential forms through $\bm{e}_A
= e_{A\mu} \bm{d}x^\mu$ (see Appendix A); this will greatly simplify
our analysis later. In defining Eq.~$(\ref{9})$, we have explicitly
assumed that the partial differential equations in Eq.~$(\ref{8})$ are
integrable. While this is not true for general spacetimes because of
the presence of singularities, they are integrable in the
weak gravity limit that we are working in. We also note that the
length of $X_a$ is proportional to the proper length $s_1$ of
$\gamma_\chi^\tau$.

Like the world function, $X_a(\tau)$ is
an extended function of two points\textemdash one on the worldline of
$\mathcal O$ and the other on the worldline of $\mathcal P$. Indeed,
it is straightforward to see that the spatial coordinates in the FWC
are an approximation of Eq.~$(\ref{9})$ by taking $\chi^\mu$ as a constant
equal to its value on $\mathit{\gamma}_\chi^\tau(0)$; the remaining
integral is proportional to the world function. We emphasize, however,
(the index $a$ notwithstanding) that $X_a$ is  the integral of a
differential form, and is a scalar function on $\mathbb{M}$ (see
Appendix A).

The construction of the time component of the test particle
$X_0(\tau)$ is more complicated because of causality. As Synge pointed
out in his derivation of the FWC, using \textit{spacelike} geodesics
$\mathit{\gamma}_\chi^\tau(s)$ in Eq.~$(\ref{9})$ is somewhat
artificial; no physical measurements ever take place along spacelike
geodesics. Strictly speaking, we should have instead used null
vectors\textemdash corresponding to optical measurements\textemdash
in the above. However, this would have resulted in a set of
optical coordinates, and since we are primarily interested in the
motion of a nonrelativistic test particle, would have been needlessly
complicated. Instead, we note that in the nonrelativistic limit the
forward lightcone of the observer opens up, and the observer's null
geodesic is well approximated by a spacelike geodesic in this
limit; we would thus expect the above construction of $X_a$ to be
valid in the nonrelativistic limit. The same argument cannot be made
for the coordinate time  $X_0$ of $\mathcal P$, however; causality
still has to be taken into account.  For this coordinate, we first
construct $X_0$ using null geodesics, and then take the appropriate
nonrelativistic limit.

Figure\ref{Fig-3} shows explicitly the null geodesic and spatial
geodesics that we use in this construction. At a time $\tau'<\tau$,
let $\mathit{\gamma}_\pi(\sigma)$ be a null geodesic that connects
$\mathcal O$ at time $\tau'$ to $\mathcal P$ at time $\tau$:
$\mathit{\gamma}_\pi(0)= \mathit{\Gamma}_u(\tau')$ and
$\mathit{\gamma}_\pi(\sigma_1)=\mathit{\Gamma}_c(\tau)$. We define
\begin{equation}
X_0(\tau) =
\int_{\mathit{\gamma}_\pi(0)}^{\mathit{\gamma}_\pi(\sigma_1)}
e_{0\mu}\bigg(\tau(\sigma), s(\sigma)\bigg)\pi^\mu(\sigma) d\sigma
-\left(-\int_{\tau'}^\tau
e_{0\mu}(\tilde{\tau},0)u^\mu(\tilde{\tau})d\tilde{\tau}\right).
\label{10}
\end{equation}
The first term in Eq.~$(\ref{10})$ is the time it takes an optical
signal to reach $\mathcal P$. The second term is the amount of time that
passes for the \textit{observer} for the optical signal to reach
$\mathcal P$. Since ${}_oe_{0\mu} = u^\mu$,
\begin{equation}
X_0(\tau) = - \left(\tau-\tau'\right) +
\int_{\mathit{\gamma}_\pi(\sigma)} \bm{e}_0.
\label{11}
\end{equation}
Note that because $u^\mu\chi_\mu=0$, unlike $X_a$ we cannot simply
replace $\pi^\mu$ by $\chi^\mu$ in the nonrelativistic limit; the
second term in Eq.~$(\ref{11})$ would vanish automatically. This
limit has to be taken much more carefully \footnote{In Fermi-Walker
coordinates $X^0 = \tau$.}.

The position of the test particle was arbitrary and could have been
placed at any point near $\mathit{\Gamma}_u(\tau)$. Thus the GLF is a
combination of a tetrad frame $\{e_A^{\>\>\>\>\mu}\}$ fixed to the
worldline of the observer together with the coordinates
$(\tau,X^a)$. It is important to note that $X^a$ measures the
\textit{relative} separation between the worldlines of $\mathcal O$
and of $\mathcal
P$. For ``small'' $X^a$ \footnote{Meaning that $\vert X^a\vert$ is
small in comparison to the scale on which the Riemann curvature tensor
varies.}, and for $\mathit{\Gamma}_u(\tau)$ and
$\mathit{\Gamma}_c(\tau)$ geodesics, $X^a$ is simply the geodesic
deviation between $\mathcal O$ and $\mathcal P$. We have also chosen,
as most physical, to use the proper time of the observer as our
time coordinate; Eq.~$(\ref{11})$ gives the dependence of the test
particle's coordinate time on $\tau$. As we usually use the frame
where the observer is at rest in the GLF, $\tau$ coincides with the
coordinate time of $\mathcal O$. We shall simply use $\tau$ as the
time variable for $\mathcal O$ to avoid introducing additional
notation.

In what follows, the final expressions of all physical quantities
measured in the GLF will be expressed in terms of $(\tau, X^a)$, the
coordinates that $\mathcal O$ measures in the GLF. In the frame,
$u^A \equiv e^A_{\>\>\>\>\mu} u^\mu = \delta^A_0$, and $u^A$ points
along the time direction for the observer, while $\chi^A(\tau,s)
\equiv e^A_{\>\>\>\>\mu} \chi^\mu(\tau,s)$ is a scalar function in the
GLF such that
\begin{equation}
\frac{\partial \chi^A}{\partial s} = \frac{D\chi^\mu}{\partial
s}e^A_{\>\>\>\>\mu} +  \chi^\mu\frac{D e^A_{\>\>\>\>\mu}}{\partial
s} = 0.
\label{12}
\end{equation}
The first term vanishes because $\mathit{\gamma}_\chi^\tau(s)$
is a geodesic, and the second term vanishes from the construction of
$e^A_{\>\>\>\>\mu}$ in Eq.~$(\ref{5})$. Thus in the GLF $\chi^A=
\chi^A(\tau)$ only. In addition, since $0=\chi_\mu u^\mu = \chi_A
\delta^A_0$, $\chi^0 = 0$ and $\chi^A$ is a unit spatial vector
pointing directly at the test particle at any time $\tau$
\footnote{Note that this only holds because we are dealing with
nonrelativistic particles. If the particle were relativistic, then we
would have to go back and use null-geodesics in constructing $X_a$.}.
As a final point on notation, although $\partial_a \equiv
\partial/\partial X^a = e_a^{\>\>\>\>\mu}\partial_\mu$, to avoid
confusion we shall always write $\partial/\partial\tau$ instead of
$\partial_0$, which is instead reserved for $\partial/\partial X^0$.

\begin{figure}
\begin{center}
\includegraphics[width=0.5\textwidth]{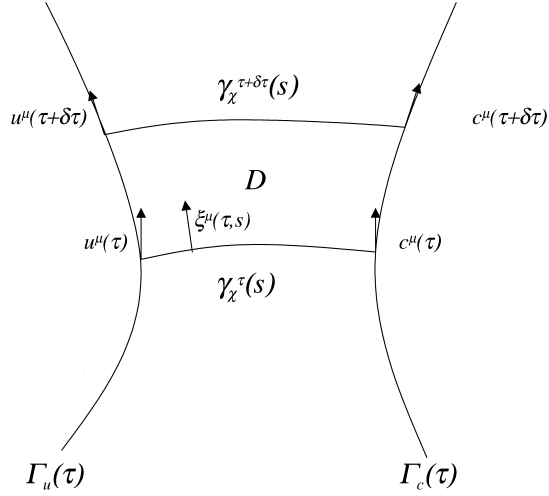}
\end{center}
\caption{\label{Fig-4} The location of $\mathcal O$ and $\mathcal P$
at two subsequent times, and the spacelike geodesics used in
constructing $X_a$ as they propagate along their worldlines. The
closed rectangular region $D$ used in Eq.~$(\ref{14})$ is shown
also. Notice that the vector $\xi^\mu$ as it varies along
$\mathit{\gamma}_\chi^\tau(s)$.}
\end{figure}

We now turn our attention to finding the 4-velocity of a test particle
as measured in the GLF. To do so we refer to Fig.~\ref{Fig-4}, which
shows the position of both the observer and the particle at two subsequent
times $\tau$ and $\tau + \delta\tau$.
Note that both the observer and the observed are moving: $\mathcal O$
along its worldline $\mathit{\Gamma}_u(\tau)$ and $\mathcal P$ along
its worldline $\mathit{\Gamma}_c(\tau)$. The observer can only measure
the relative $4$-velocity between $\mathcal P$ and himself. Beginning
with the spatial coordinates, we take
\begin{eqnarray}
\frac{d X_a}{d \tau} &\equiv& \lim_{\delta\tau\to 0}
                 \frac{X_a(\tau+\delta\tau)-X_a(\tau)}
                 {\delta\tau},
\nonumber \\
&=& \lim_{\delta\tau\to 0} \frac{1}{\delta\tau}
      \left\{
      \int_{\mathit{\gamma}_\chi^{\tau+\delta\tau}(s)} \bm{e}_a
      -
      \int_{\mathit{\gamma}_\chi^\tau(s)} \bm{e}_a
      \right\},
\label{13}
\end{eqnarray}
where $\mathit{\gamma}_\chi^{\tau+\delta\tau}(s)$ and
$\mathit{\gamma}_\chi^{\tau}(s)$ are spacelike geodesics from $\mathcal O$
to $\mathcal P$ at $\tau +\delta\tau$ and $\tau$, respectively (see
Fig. 3).  Adding and subtracting integrals along the worldlines of
$\mathcal O$ and $\mathcal P$,
\begin{eqnarray}
\frac{d X_a}{d \tau} &=& -\lim_{\delta\tau\to 0} \frac{1}{\delta\tau}
      \left\{\int_{\partial D}\bm{e}_a
      + \int_\tau^{\tau+\delta\tau}e_{a\mu}(\tilde{\tau},0)u^\mu(\tilde\tau)
      d\tilde\tau
      - \int_\tau^{\tau+\delta\tau}e_{a\mu}(\tilde{\tau},s_1)c^\mu(\tilde{\tau})
      d\tilde\tau\right\},
\nonumber \\
&=& c_a(\tau,X^a)- \lim_{\delta\tau\to 0}
      \frac{1}{\delta\tau}\int_D\bm{de}_a,
\label{14}
\end{eqnarray}
where $c_A = e_{A\mu}(\tau, X^a) c^\mu(\tau)$ is the $4$-velocity of
the test particle in the GLF. $D$ is the closed region bounded by
$\mathit{\gamma}_\chi^\tau(s)$,
$\mathit{\gamma}_\chi^{\tau+\delta\tau}(s)$, and the worldlines
$\mathit{\Gamma}_u(\tau)$ and $\mathit{\Gamma}_c(\tau)$ between $\tau$
and $\tau +\delta\tau$.

We parameterize the region $D$ by the $1$-forms $\bm{d}l^\mu_1 =
\chi^\mu\bm{d}s$ and $\bm{d}l^\mu_2 = \xi^\mu\bm{d}\tau$ where
\begin{equation}
\xi^\mu \equiv \frac{D \chi^\mu}{\partial\tau},
\label{15}
\end{equation}
is a function of $(\tau,s)$. Because $\chi_\mu\chi^\mu = 1$, then
$\xi_\mu\chi^\mu = 0$, and $\bm{d}l^\mu_1$ and $\bm{d}l^\mu_2$ are
linearly independent. From Eq.~$(\ref{64})$, $\bm{de}_a =
\bm{e}^B\wedge\bm{\omega}_{aB}$ where $\bm{\omega}_{aB} =
e_a^{\>\>\>\>\nu}\nabla_\mu e_{B\nu} \bm{d}x^\mu$ is the Ricci
coefficient 1-form. Then,
\begin{equation}
\int_D \bm{de}_a = -\int_\tau^{\tau +\delta\tau}\int_0^{s_1}
            \left(
            e^B_{\>\>\>\>\mu}\omega_{\nu aB}
                -
            e^B_{\>\>\>\>\nu}\omega_{\mu aB}
            \right)
            \xi^\mu\chi^\nu d\tilde\tau ds.
\label{16}
\end{equation}
The limit $\delta\tau\to 0$ is now trivial to take, leaving only a path
integral along $\bm{\gamma}_\chi^\tau(s)$. Then
\begin{eqnarray}
c_a &=& \frac{d X_a}{d\tau}+\int_0^{X_a}
    \left\{
        \omega_{Bac}-\omega_{caB}
    \right\} \xi^Bd\tilde{X}^c,
\nonumber \\
   &=& \frac{d X_a}{d\tau}+\int_0^{X_a}
    \left\{
        \langle{\xi, \bm{\omega}_{ac}}\rangle \bm{e}^c
        -
                  \langle{\xi, \bm{e}^B}\rangle\bm{\omega}_{aB}
    \right\},
\label{17}
\end{eqnarray}
where we have now expressed the path integral in GLF coordinates by using
$d\tilde{X}^c = e^c_{\>\>\>\>\mu}\>\chi^\mu d\tilde{\tau}$, and
$\xi^A\equiv e^A_{\>\>\>\>\mu}\xi^\mu$. As usual, $\langle{\xi,
\bm{\omega}_{ac}}\rangle = \xi^\mu \omega_{\mu ac}$.

To determine $\xi$, we note that $\xi^\mu$ measures the deviation in
the geodesic $\mathit{\gamma}_\chi^{\tau+\delta\tau}(s)$ from
$\mathit{\gamma}_\chi^{\tau}(s)$ at any time $\tau$ along $s$, and is
thus the solution of the geodesic deviation (or Jacobi) equation in $s$
\cite{Synge}. In the GLF,
\begin{equation}
0=\frac{\partial^2\xi^A}{\partial s^2} + \tilde{R}^A_{\>\>\>\>B}\xi^B
\label{18}
\end{equation}
where $\xi^A\equiv e^A_{\>\>\>\>\mu}\xi^\mu$ and
$\tilde{R}^A_{\>\>\>\>B}\equiv R^A_{\>\>\>\>CBD}\chi^C\chi^D$ (see
Appendix A). Note that $\chi^A$ is independent of $s$ and only fixes a
direction in the above.

 From Fig.~\ref{Fig-4} we see that $\xi^A$ interpolates between the tangent
vector to the worldline of $\mathcal O$, and the tangent vector to the
worldline of $\mathcal P$. Since $\xi^A\chi_A=\xi^\mu\chi_\mu=0$, the
precise boundary conditions are: $\lim_{s\to0^{+}} \xi^A(\tau,s)=
\delta^A_0$ and $\lim_{s\to s_1^{-}} \xi^A(\tau,s) = \left(\delta^A_B -
\chi^A\chi_B\right)c^B$. However, $\xi^A$ appears in Eq.~$(\ref{16})$
through of the 2-form  $\bm{d}l^\mu_1\wedge\bm{d}l^\nu_2$. Since
$\bm{d}l^\mu_2$ lies parallel to $\chi^\mu$, this automatically
projects to zero any component of $\xi^A$ parallel to
$\chi^A$. Consequently, we can without loss of generality replace the
second boundary condition by the much simpler condition $\lim_{s\to
s_1^{-}} \xi^A(\tau,s) = c^A$. Equation $(\ref{18})$ can then be solved
iteratively, and
\begin{equation}
\xi^A(\tau,s) = \frac{1}{s_1}\left\{s\> c^A(\tau)+(s_1-s)\delta^A_0\right\}+
\int_0^{s_1} \tilde{R}^A_{\>\>\>\>B}(\tau,s') \xi^B(\tau,s')
G(s,s')ds',
\label{19}
\end{equation}
where $G(s,s')$ is the Green's function
\begin{equation}
G(s,s') =\frac{s}{s_1}(s_1-s')\theta(s'-s)+
        \frac{s'}{s_1}(s_1-s)\theta(s-s'),
\label{20}
\end{equation}
and $\theta(x)$ is the Heaviside function.

\begin{figure}
\begin{center}
\includegraphics[width=0.5\textwidth]{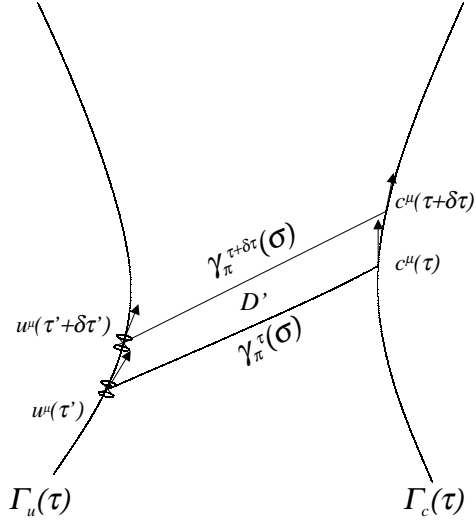}
\end{center}
\caption{\label{Fig-5} The null geodesics used in constructing $X_0$
at two subsequent times as $\mathcal O$ and $\mathcal P$ propagate
along their worldlines. The closed region $D'$ used in
Eq.~$(\ref{21})$ is shown.}
\end{figure}

For $c_0$ we proceed the same way. Using now the diagram in
Fig.~\ref{Fig-5},
\begin{equation}
c_0=\frac{dX_0}{d\tau} + \lim_{\delta\tau\to0}
      \frac{1}{\delta\tau}\int_{D'}\bm{e}^a\wedge\bm{\omega}_{0a}.
\label{21}
\end{equation}
To parameterize $D'$, we again take $\bm{d}l^\mu_1 =
\pi^\mu\bm{d}\sigma$, but now $\bm{d}l^\mu_2 =
\zeta^\mu\bm{d}\tau$ where, in the GLF,
\begin{equation}
\zeta^A(\tau,\sigma) = \delta^A_0 +
\left\{c^A(\tau)-\delta^A_0\right\}\frac{\sigma}{\sigma_1}.
\label{22}
\end{equation}
Clearly, $\zeta^A$ and $\pi^A$ are linearly independent. Unlike the
spatial components of the 4-velocity, we \textit{cannot} use
$e^A_{\>\>\>\>\mu}D\pi^\mu/\partial\tau$ to parameterize $D'$, because
it is either a null vector or a spacelike vector. Although like
$\xi^A$, $\zeta^A$ interpolates between the tangent vector at
$\mathit{\Gamma}_u(\tau')$ and the tangent vector at
$\mathit{\Gamma}_c(\tau)$, it does \textit{not} include the
corrections due to nonvanishing curvature that Eq.~$(\ref{18})$
does. Because we will be working in the linearized gravity limit, such
correction terms are of the order $h_{AB}$ times the particle
velocity, and can be neglected to lowest order. We therefore get
\begin{equation}
c_0=\frac{dX_0}{d\tau} + \int_{\bm{\gamma}_\pi^\tau(\sigma)}
    \left\{
        \langle\zeta,\bm{\omega}_{0A}\rangle \bm{e}^A
        -
        \langle\zeta,\bm{e}^A\rangle \bm{\omega}_{0A}
    \right\}.
\label{23}
\end{equation}

Equations $(\ref{17})$ and $(\ref{23})$ in principle determine the
components of the $4$-velocity $c^A$ of the test particle in the
GLF. However, because both $\xi^A$ and $\zeta^A$ themselves depend
$c^A$, these equations form a set of coupled integral equations in
$c^A$. While these equations can be solved iteratively using
Eq.~$(\ref{19})$ and Eq.~$(\ref{22})$, we are primarily interested in
the behavior of nonrelativistic test particles in the linearized
gravity limit. In fact, much of our construction of the coordinates
for the GLF is only valid for the nonrelativistic test particle. We thus keep
only terms linear in $h_{\mu\nu}$ where $g_{\mu\nu} =  \eta_{\mu\nu} +
h_{\mu\nu}$ (see Appendix A), and we approximate $c^0 \approx 1$,
keeping only terms linear in $c^A-\delta^A_0$. Since both the spatial
velocity and the curvature effects are small, we also neglect cross
terms of $O((c^A-\delta^A_0)h_{AB})$. Thus, we can neglect curvature
corrections in Eq.~$(\ref{19})$ altogether, and can take $\xi^A \approx
\delta^A_0$ since $\xi^A$ always appears in the combination
$\xi\bm{\omega}$. Using a similar argument, we take
$\zeta^A\approx\delta^A_0$ as well [which is why we did not have to be
concerned with curvature corrections to $\zeta^A$ in
Eq.~$(\ref{22})$]. With these approximations,
\begin{equation}
c_A = \frac{dX_A}{d\tau} +
\int_0^{X^a}\left\{\omega_{0Ab}-\omega_{bA0}\right\} d\tilde{X}^b,
\label{24}
\end{equation}
where $d\tilde{X}^B = e^B_{\>\>\mu}\chi^\mu d\tau$.

\section{The Hamiltonian}

The action for a test particle with mass $m$ and charge $q$
in general coordinates is
\begin{equation}
I = -m\int\sqrt{-c_\mu c^\mu}d\tau + q\int A_\mu c^\mu d\tau.
\label{25}
\end{equation}
Because Eq.~$(\ref{25})$ is time reparameterization invariant, we are
free to choose the proper time $\tau$ of $\mathcal O$ as our
parameterization. In the GLF the Lagrangian becomes
\begin{eqnarray}
{\mathcal L}&=& -m\sqrt{(c^0)^2 - (c^a)^2} + qA_0c^0 + qA_ac^a,
\nonumber \\
    &\approx& -m + \frac{1}{2} m\left(\frac{dX^a}{d\tau}+\int_0^{X^a}
        \left\{
            \omega_0^{\>\>a}{}_b-\omega_b^{\>\>a}{}_0
        \right\}
        d\tilde{X}^b\right)^2 - m\int_0^{X^a}\omega_{00b}d\tilde{X}^b
\nonumber \\
&{}& + qA_0 \left(1+\int_0^{X^a}\omega_{00b}d\tilde{X}^b\right) +
qA_a\left(\frac{dX^a}{d\tau}+\int_0^{X^a}
        \left\{
            \omega_0^{\>\>a}{}_b-\omega_b^{\>\>a}{}_0
        \right\}
        d\tilde{X}^b
        \right),
\label{26}
\end{eqnarray}
where $A_B\equiv e_B^{\>\>\>\>\mu}A_\mu$ is the vector potential in
the GLF. In the above we have used the nonrelativistic and
linearized gravity limits. This includes taking $dX^0/d\tau\approx
1$. The momentum canonical to $X^a$ is then:
\begin{equation}
p_a\equiv \frac{\delta{\mathcal L}}{\delta (dX^a/d\tau)} =
m\frac{dX_a}{d\tau} + m\int_0^{X^a}
\left\{\omega_{0ab} -\omega_{ba0}\right\}d\tilde{X}^b + qA_a,
\label{27}
\end{equation}
so that in general
\begin{eqnarray}
H =&{}& \frac{1}{2m}\left(p_a - qA_a\right)^2 - p_a
    \int_0^{X^a}
    \left\{
        \omega_0^{\>\>a}{}_b-\omega_b^{\>\>a}{}_0
    \right\}d\tilde{X}^b
\nonumber \\
&{}&    -qA_0\left(1+\int_0^{X^a}\omega_{00b}d\tilde{X}^b\right) +
    m\int_0^{X^a}\omega_{00b}d\tilde{X}^b.
\label{28}
\end{eqnarray}
There are two special cases to consider.

\subsection{Stationary metrics}

For stationary metrics there is always a frame where $g_{\mu\nu}$
is independent of time. In this frame,
\begin{equation}
\omega_{00b}=\frac{1}{2}\partial_b h_{00}, \qquad
\omega_{0ab}-\omega_{ba0} = \frac{1}{2}\partial_b h_{a0},
\label{29}
\end{equation}
where we have used Eq.~$(\ref{70})$. Then,
\begin{eqnarray}
c_0 &=& \frac{dX_0}{d\tau} +\frac{1}{2}\left\{h_{00}(\tau,X^a)-
h_{00}(\tau,0)\right\},
\nonumber \\
c_a &=& \frac{dX_a}{d\tau} +\frac{1}{2}\left\{h_{a0}(\tau, X^a)-
h_{a0}(\tau, 0)\right\},
\label{30}
\end{eqnarray}
while
\begin{eqnarray}
H = &{}&\frac{1}{2m}(p_a-qA_a)^2 -
\frac{1}{2}p_a\left\{h^a_0(\tau, X^a)-h^a_0(\tau, 0)\right\}
+\frac{1}{2}m\left\{h_{00}(\tau, X^a)-h_{00}(\tau, 0)\right\}
\nonumber \\
&{}&- qA_0\left\{1+h_{00}(\tau, X^a)-h_{00}(\tau, 0)\right\}.
\label{31}
\end{eqnarray}
To compare this with DeWitt's results \cite{deW1}, \cite{deW2}, we
first remember that we are working in the (orthogonal) GLF, while DeWitt
is working in the general coordinate frame. To transform the above
back to the general coordinate frame, we use Eq.~$(\ref{69})$ in
\begin{equation}
p_a =\left(\delta_a^\mu - \frac{1}{2}h_a^\mu\right)p_\mu\approx
p_{\hat{a}} -\frac{m}{2}h_{\hat{a}\hat{0}},
\label{32}
\end{equation}
where we follow DeWitt and neglect terms of $O(ph)$. For clarity, we
use a caret to distinguish between indices in the GLF and DeWitt's
frame. Like DeWitt we neglect terms $O(Ah)$ as well, and find
\begin{equation}
H=
\frac{1}{2m}\left(p_{\hat{a}}-mh_{\hat{a}\hat{0}}(\tau, X^{\hat{a}})-
   \frac{m}{2}h_{\hat{a}\hat{0}}(\tau, 0)-qA_{\hat{a}}\right)^2
+ \frac{m}{2}\left\{h_{\hat{0}\hat{0}}(\tau, X^{\hat{a}})-
h_{\hat{0}\hat{0}}(\tau, 0)\right\} +
qA_{\hat{0}}.
\label{33}
\end{equation}
This agrees with DeWitt's result up to a constant shift in velocity and
in energy. This shift is needed because DeWitt's coordinates are fixed
to the origin of the mass generating the gravitational field (for
example, on the center of the Earth), while our origin is fixed on the
observer (on the surface of the Earth, say).

\subsection{For gravitational waves}

For GWs we work in the usual TT gauge. then $\omega_{00b}=0$ now, while
\begin{equation}
\omega_{0ab}-\omega_{ba0} = -\frac{1}{2}\frac{\partial h_{ab}}{\partial
\tau},
\label{34}
\end{equation}
from which we see that
\begin{equation}
c_0 = \frac{dX_0}{d\tau}, \qquad c_a = \frac{dX_a}{d\tau} -
N_a(X^a), \label{35}
\end{equation}
where
\begin{equation}
N_a(X^a) = \int_0^{X^a} \bm{\omega}_{a0}.
\label{36}
\end{equation}
Then,
\begin{equation}
H = \frac{1}{2m}(p_a-qA_a)^2 + p_aN^a - qA_0.
\label{37}
\end{equation}
Now, in the long-wavelength limit, $N_a=X^b(\partial
h_{ab}/\partial\tau)/2$. Thus, for neutral particles Eq.~$(\ref{37})$
agrees with the Hamiltonian derived in \cite{ADS1995} from the geodesic
deviation equation. In that paper one of us (A.D.S.) also postulated the
existence of a minimal coupling with the  EM field which was of the
form $\sim(p_a+N_a-qA_a)^2$ using the notation of this paper. This was
based on the usual arguments for minimal coupling, and was valid within
the framework and approximations of that paper. This direct coupling
between $N_a$ and $A_a$ does not appear in this more general
analysis. However, a direct coupling betwen the spin connection and
the EM vector potential is present for fermionic systems as presented
in \cite{Chiao_0}.

\section{Properties of $N_a$}

If we expand out the kinetic part of the Hamiltonian Eq.~$(\ref{37})$,
we see that for small $A_a$ both $A_a $ and $N_a$ couple to the test
particle in much the same way. Surprisingly, this similarity between $N_a$
and $A_a$ goes much deeper. Like $A_a$, $N_a$\textemdash which has
units of velocity\textemdash is transverse and satisfies the wave
equation in GLF. Like $A_a$, $N_a$ is a gauge dependent object, the
gauge symmetry in this case being the choice of local Lorentz frames,
and the gauge group being the Galilean group (since we are dealing with
nonrelativistic test particles). As with $N_a$, we can construct from
$N_a$ effective ``electric'' and ``magnetic'' fields that obey partial
differential equations that have the same form as Maxwell's
equations. These fields have direct physical meaning, and
they turn out to be proportional to the integrals of the electric and
magnetic parts of the Weyl tensor.

Consider first
\begin{eqnarray}
\frac{\partial N_a}{\partial\tau} &=& \lim_{\delta\tau\to0}
            \frac{1}{\delta\tau}
            \left\{
                N_a(\tau+\delta\tau, X^a)
                -
                N_a(\tau,X^a)
            \right\},
\nonumber \\
&=& \omega_{Bao}(\tau,X^a)c^B(\tau) -
    \omega_{B a0}(\tau,0)\delta^B_0
    -\lim_{\delta\tau\to0} \frac{1}{\delta\tau}\int_D \bm{d\omega}_{a0},
\label{38}
\end{eqnarray}
where we have used the same arguments as in Sec.~II. Once again,
because we are in the nonrelativistic limit, the domain $D'$ in
Fig.~\ref{Fig-5} is well approximated by the domain $D$ in
Fig.~\ref{Fig-4}. Using Eq.~$(\ref{65})$ and the usual
parameterization of $D$,
\begin{eqnarray}
\frac{\partial N_a}{\partial\tau} &=& -\frac{1}{2} \lim_{\delta\tau\to0}
            \frac{1}{\delta\tau}\int_D
            R_{BCao}\bm{d}l_1^B\wedge\bm{d}l_2^C,
\nonumber \\
&=& -\int_0^{X^a} R_{b0a0}d\tilde{X}^b.
\label{39}
\end{eqnarray}
since $c^A\approx \delta^A_0$ and $\omega_{0a0}=0$ for GWs in the TT
gauge. Similarly,
\begin{eqnarray}
\frac{\partial^2 N_a}{\partial\tau^2} &=& \lim_{\delta\tau\to0}
\frac{1}{\delta\tau}\int_D \partial_C R_{B0a0}\bm{d}l_1^C\wedge\bm{d}l_2^B,
\nonumber \\
&=& -\int_0^{X^a}\partial_0 R_{b0a0} d\tilde{X}^b,
\label{40}
\end{eqnarray}
since $R_{00b0}=0$.

\begin{figure}
\begin{center}
\includegraphics[width=0.5\textwidth]{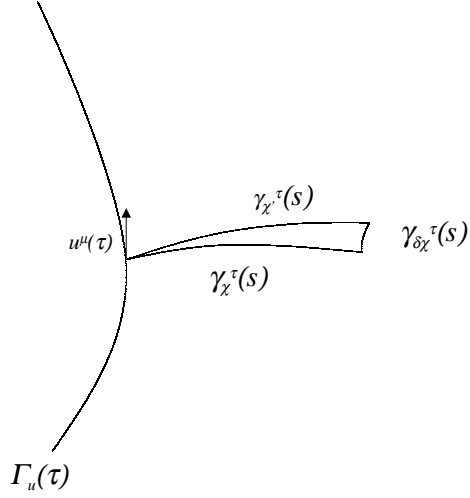}
\end{center}
\caption{\label{Fig-6}A sketch of the two spacelike geodesics
$\mathit{\gamma}_\chi^\tau(s)$ and $\mathit{\gamma}_{\chi'}^\tau(s)$
for $X^a$ and $X^a +\delta X^b$ respectively used in calculating
spatial derivatives. The addition of
$\mathit{\gamma}_{\delta\chi}^\tau(s)$ linking the two points defines
the closed triangular region $\mathit\Delta$ used in Eq.$(\ref{41})$.}
\end{figure}

For the spatial derivatives, we refer to Fig.~\ref{Fig-6} and now
consider the three spacelike geodesics bounding the closed surface
$\mathit{\Delta}$ formed by $\mathit{\gamma}_\chi^\tau(s)$, connecting
the origin to $X^a$; $\mathit{\gamma}_{\chi'}^\tau(s)$, 
connecting the origin to $X^a+\delta X^b$ for $\delta X^b$ small; and
$\mathit{\gamma}_{\delta\chi}^\tau(s)$, connecting $X^a$ to $X^a +
\delta X^b$.
Then,
\begin{eqnarray}
\frac{\partial N_a}{\partial X^b} &=& \lim_{\delta X^b\to0}\frac{1}{\delta X^b}
        \left\{
        N_a(\tau,X^a+\delta X^b) - N_a(\tau, X^a)
        \right\},
\nonumber \\
&=& \lim_{\delta X^b\to0} \frac{1}{\delta X^b}
        \left(
            \int_{\mathit{\gamma}_{\delta\chi}^\tau(s)}
                \bm{\omega}_{a0}
            -
            \int_{\mathit{\Delta}}\bm{d\omega}_{a0}
        \right),
\label{41}
\end{eqnarray}
where we have once again added and subtracted an integral\textemdash
along $\mathit{\gamma}_{\chi_\Delta}^\tau(s)$\textemdash and used
Stokes's theorem on the closed boundary $\partial\mathit{\Delta}$ of
$\mathit\Delta$. Parameterizing $\mathit\Delta$ by $\bm{d}l_1^\mu =
\chi^\mu\bm{d}s$ and $\bm{d}l_2^\mu ={\chi'}^\mu\bm{d}s'$, then
\begin{equation}
\partial_b N_a = \omega_{ba0} - \int_0^{X^a}
   R_{cba0}d\tilde{X}^c.
   \label{42}
\end{equation}
This equation shows explicitly the difference between taking spatial
derivatives in flat spacetime verses taking derivatives in curved
space. In flat spacetimes the gradient of $N_a$ would simply be
$\omega_{ba0}$. In curved space, on the other hand, we have to take
into account the differences in curvature between $X^a$ and $X^a +
\delta X^b$, and that introduces the additional curvature
term. Proceeding similarly,
\begin{eqnarray}
\partial_b\partial_c N_a & = & \partial_c
\omega_{ba0} -  R_{cba0}(\tau, X^a) + \lim_{\delta X^c\to0}
\frac{1}{\delta X^c} \int_{\mathit\Delta}
\partial_e R_{dba0}\bm{d}l^e_1\wedge \bm{d}l^d_2,
\nonumber \\
&=& \partial_c\omega_{ba0} -R_{cba0}(\tau,0) -\int_0^{X^a} \partial_c
R_{eba0}d\tilde{X}^e.
   \label{43}
\end{eqnarray}

From Eq.~$(\ref{42})$,
\begin{equation}
\partial^a N_a  = \omega^a_{\>\>\>a0} + \int_0^{X^a}R_{a0}d\tilde{X}^a.
\label{44}
\end{equation}
For GWs in the TT gauge $\omega^a_{\>\>\>a0} = {(\partial
h^a_a/\partial\tau})/2 =0$ and $R_{a0} = -\Box h_{a0}/2
= 0$, where $\Box$ is the d'Alembertian operator. $N_a$ is thus
transverse, and $\partial^a N_a =0$. Next, from Eqs.~$(\ref{43})$ and
$(\ref{40})$,
\begin{equation}
\Box N_a = -\int_0^{X^a} \left(\frac{\partial
R_{ab}}{\partial\tau}\right) d\tilde{X}^b.
\label{45}
\end{equation}
But this also vanishes since $R_{ab}=0$. Thus, like the vector
potential for EM, $N_a$ is a transverse vector field obeying the wave
equation $\Box N_a = 0$.

Next, to construct the effective ``electric'' and ``magnetic'' fields
from $N_a$, we consider the four-vector $N_A$, constructed from $N_a$
with $N_0\equiv 0$, much like the vector potential for EM in the
 Coulomb gauge. Then clearly $\partial^A N_A=0$, and $N_A$ is a
transverse field. We then define an effective field strength
$\mathfrak{F}_{AB} = \partial_A N_B - \partial_B N_A$ with the
corresponding ``electric'' and ``magnetic'' fields
\begin{eqnarray}
\mathfrak{E}_a &\equiv& \mathfrak{F}_{a0} =
        -\frac{\partial N_a}{\partial\tau},
\nonumber \\
\mathfrak{B}^a &\equiv& \frac{1}{2} \epsilon^{abc}\mathfrak{F}_{bc} =
        \epsilon^{abc} \partial_b N_c.
\label{47}
\end{eqnarray}
Using the same methods as above, we see that
\begin{eqnarray}
\frac{\partial\mathfrak{E}_a}{\partial\tau} &=& \int_0^{X^a}
\left(\frac{\partial R_{b0a0}}{\partial\tau}\right) d\tilde{X}^b,
\nonumber \\
\partial_b\mathfrak{E}_a &=& R_{b0a0}(\tau,0) + \int_0^{X^a}
\partial_b R_{c0a0} d\tilde{X}^c,
\nonumber \\
\frac{\partial\mathfrak{B}^a}{\partial\tau} &=& -\int_0^{X^a}\epsilon^{abc}
\left(\frac{\partial R_{ebc0}}{\partial\tau}\right) d\tilde{X}^e,
\nonumber \\
\partial_b\mathfrak{B}^a &=&
-\epsilon^{afc} R_{bfc0}(\tau,0)
-\int_0^{X^a}\epsilon^{afc}
\partial_b R_{efc0} d\tilde{X}^e,
\label{48}
\end{eqnarray}

Now, it is clear that by definition
$\epsilon^{ABCD}\partial_B\mathfrak{F}_{CD}=0$, and this relation
must explicitly hold if $\mathfrak{F}_{AB}$ has been defined
consistently. To verify this, we first consider
\begin{equation}
\partial^a {\mathfrak{B}_a} = -\epsilon^{afc}
R_{afc0}(\tau,0)-\int_0^{X^a}\epsilon^{afc}
\partial_a R_{efc0} d\tilde{X}^e = 0,
\label{49}
\end{equation}
which vanishes identically using the 1st (Eq.~$(\ref{66})$) and 2nd
Bianchi Identities (Eq.~$(\ref{67})$). Similarly,
\begin{equation}
\frac{\partial \mathfrak{B}^a}{\partial\tau}
-\epsilon^{abc}\partial_b\mathfrak{E}_c
=\epsilon^{abc}\int_0^{X^a}d\tilde{X}^e\left(-\frac{\partial
R_{ebc0}}{\partial\tau} + \partial_b R_{e0c0}\right),
\label{50}
\end{equation}
also vanishes for the same reason. Thus, $\epsilon^{ABCD}\partial_B
\mathfrak{F}_{CD}\equiv 0$ is as a direct consequence of
the Bianchi Identities for the Riemann curvature tensor.

For the other part of Maxwell's equations,
$\partial^A\mathfrak{F}_{AB}=0$, we consider
\begin{equation}
\partial^a \mathfrak{E}_a = -R_{00}(\tau, 0) +
\int_0^{X^a}\partial^a R_{c0a0}d\tilde{X}^c.
\label{51}
\end{equation}
From Eq.~$(\ref{67})$,
\begin{equation}
0=\partial^A R_{C00A} +\partial_c R_{00} - \frac{\partial
R_{0C}}{\partial\tau}.
\label{52}
\end{equation}
Once again, $R_{AB}=0$ implies that
$\partial^a\mathfrak{E}_a=0$. Finally, we
consider
\begin{equation}
\epsilon^{abc}
\partial_b\mathfrak{B}_c +
\frac{\partial\mathfrak{E}^a}{\partial\tau} = -\int_0^{X^a}
\left(\frac{\partial R^a_{\>\>\>b}}{\partial\tau}\right) d\tilde{X}^b = 0,
\label{53}
\end{equation}
using Eq.~$(\ref{67})$ and $R_{ab}=0$ once again. Consequently,
$\mathfrak{E}_a$ and $\mathfrak{B}^a$ satisfy equations that have the
same form as Maxwell's equations. Notice also that like EM waves,
$\partial^A\mathfrak{F}_{AB}=0$ only in the absence of sources.

The surprising connection between $\mathfrak{E}_a$ and
$\mathfrak{B}^a$, and Maxwell's equations can be understood by looking
at the ``electric'' and ``magnetic'' parts of the Weyl tensor
$C_{ABCD}$. For GWs propagating in a flat background, $C_{ABCD} =
R_{ABCD}$ and the Riemann curvature tensor can be separated into two
parts \cite{deF}: The electric part,
\begin{equation}
E_{ab} \equiv -C_{0a0b} = -R_{0a0b},
\label{54}
\end{equation}
and the magnetic part,
\begin{equation}
H^a_{\>\>b} = \frac{1}{2} \epsilon^{0aef}C_{ef0b}= -\frac{1}{2}
\epsilon^{aef}R_{efb0}.
\label{55}
\end{equation}
Clearly
\begin{equation}
\mathfrak{E}_a = -\int_0^{X^a} E_{ab}d\tilde{X}^b, \qquad
\mathfrak{B}^a = -\int_0^{X^a} H^a_b d\tilde{X}^b.
\label{56}
\end{equation}
$\mathfrak{E}_a$ and $\mathfrak{B}^a$ are simply path integrals of
$E_{ab}$ and $H^a_{\>\>b}$; they obey Maxwell's equations because
$E_{ab}$ and $H^a_{\>\>b}$ obey tensor Maxwell-like equations
\cite{deF}.

If $N_a$ functions as an effective vector potential for the GW, what,
then, is the corresponding gauge group? Notice that although we have
chosen a specific coordinate system\textemdash and thus broken general
coordinate invariance\textemdash we still have a residual invariance
left over. To see this, let us do a \textit{local} Lorentz
transformation $L_A^{\>\>\>\>\tilde{A}}$ on $e_a^{\>\>\>\>\mu}$ such
that  $\eta_{AB} = L_A^{\>\>\>\>\tilde{A}}L_{B\tilde{A}}$ while
$e_A^{\>\>\>\>\mu} =
L_A^{\>\>\>\>\tilde{A}}e_{\tilde{A}}^{\>\>\>\>\mu}$. This leaves
$g_{\mu\nu}= e_{A\mu}\>\>e^A_{\>\>\>\>\nu}=e_{\tilde{A}\mu}\>\>
e^{\tilde{A}}_{\>\>\>\>\nu}$ invariant, and we see that local Lorentz
invariance is still left over. Now, it is straightforward to show that
under a local Lorentz transformation $\omega_{CAB} \to
\omega_{\tilde{C}\tilde{A}\tilde{B}}
+  L^D_{\tilde{B}}
\partial_{\tilde{C}} L_{D\tilde{A}}$; $\omega_{CAB}$ transforms
anomalously under a local Lorentz transformation. Consequently,
\begin{equation}
N_a \to N_{\tilde{a}} + \int_0^{X^a}L^D_{\>\>\>\>\tilde{0}}
\partial_{\tilde{c}} L_{D\tilde{a}}d\tilde{X}^{\tilde{c}}.
\label{57}
\end{equation}
Because we are working in the nonrelativistic limit, the gauge
group for $N_a$ is the local Galilean group. Indeed, for a pure boost
in the nonrelativistic limit, $L_{A\tilde{A}} \approx
\delta_{A\tilde{A}} - v_aK^a_{A\tilde{A}}$ where $K^a_{A\tilde{A}}$ is
the generator of boosts. Then,
\begin{equation}
N_a \to N_{\tilde{a}} - \int_0^{X^a}dv_{\tilde{a}} = N_{\tilde{a}} -
\left\{v_{\tilde{a}}(\tau, X^a) -  v_{\tilde{a}}(\tau, 0)\right\},
\label{58}
\end{equation}
and $N_a$ changes by a local velocity field.

Finally, we calculate the equations of motion for
the test particle in a GW. Equation $(\ref{26})$ for GWs is
\begin{equation}
{\mathcal L}= \frac{m}{2}\left(\frac{dX^a}{d\tau} - N^a\right)^2 + qA_0
+qA_a\left(\frac{dX^a}{d\tau} - N^a\right),
\label{59}
\end{equation}
from which we get
\begin{equation}
m\frac{d^2X^a}{d\tau^2} = -m\left\{\mathfrak{E}^a
+\epsilon^{abc}\frac{dX_b}{d\tau}\mathfrak{B}_c\right\} + q
\left\{E^a + \epsilon^{abc}\frac{dX_b}{d\tau}B_c\right\},
\label{60}
\end{equation}
after using Eq.~$(\ref{47})$; like DeWitt we drop terms $O(Ah)$.
In Section II we stated that we would neglect terms of order
$(c^A-\delta^A_0)h_{ab}$, and yet Eq.~$(\ref{60})$ contains just such
a term. It is, however, straightforward, though tedious, to repeat the
calculation for Eq.~$(\ref{60})$ keeping the next higher order
velocity-connection terms. After doing so we still obtain the above to
lowest order.

Let us suppose that a GW can be approximated as a plane wave with
wave vector $k_a$. Notice that the integral in Eq.~$(\ref{26})$ is
always over the plane perpendicular to $k_a$, yet $h_{ab}$ is a
only function of the spatial coordinates parallel to
$k_a$. Consequently, $N_a =
\omega_{ba0}X^b$, $\mathfrak{E}_a = R_{b0a0}X^b$ and $\mathfrak{B}^a =
\epsilon^{abc}R_{bce0}X^e/2$. Thus, for planar GWs,
\begin{equation}
m\frac{d^2X_a}{d\tau^2} = -m\left\{R_{b0a0}X^b
+\frac{dX^b}{d\tau}X^cR_{abc0}\right\} + q
\left\{E_a + \epsilon_{abc}\frac{dX^b}{d\tau}B^c\right\}.
\label{61}
\end{equation}
As expected, in the low velocity limit where the $\mathfrak{B}^a$ term
can be neglected, this is just the geodesic deviation equation for a
charged test particle. Note also that Eq.~$(\ref{61})$ holds for
general GWs in the long-wavelength limit as well, and that it agrees
with the equation of motion for a charged test particle interacting
with a GW found in \cite{MTW}.

\section{Concluding Remarks}

In Eqs.~$(\ref{26})$ and $(\ref{28})$ we can clearly see the natural
separation that occurs in the dynamics of test particles in two
extremes. In the one extreme the metric for $\mathbb{M}$ can be
approximated as being stationary. Matter moving with a characteristic
velocity $v$ much smaller than the speed of light ($v/c<<1$)
makes up the dominant contribution to the stress-energy tensor in these
spacetimes; thus to a very good approximation the presence of GWs can
be neglected. In this extreme our Hamiltonian reduces to DeWitt's
Hamiltonian, and the straightforward approach to the derivation of
$H_{SF}$ described in the Introduction works. In the other extreme,
all the curvature effects are due to GWs, and the characteristic
velocity for the metric is the speed of light. In this extreme, our
Hamiltonian reduces to the Hamiltonian derived in \cite{ADS1995} for
geodesic \textit{deviation} motion in the presence of GW's in the
long-wavelength limit, and the equations of motion reduce to the usual
\textit{geodesic} deviation equations of \cite{MTW}.

Let us be very clear. The dynamics of nonrelativistic, classical test
particles in stationary spacetimes that we have derived in the GLF and
are the same as the dynamics for these particles derived using standard
methods. Thus, such classical tests of GR as the perihelion of Mercury
and the gravitational redshift will follow through in the GLF as
well. By extension, because QFCS's are usually formulated in
stationary spacetimes, we would expect such an analysis as the
evaporation of black holes and black hole thermodynamics to hold
also. It is when we consider nonstationary spacetimes\textemdash such
as those where GWs play a dominant role in determining the
physics\textemdash that we must take care to explicitly include the
observer and his experimental apparatus in the analysis.

\begin{figure}
\begin{center}
\includegraphics[width=1\textwidth]{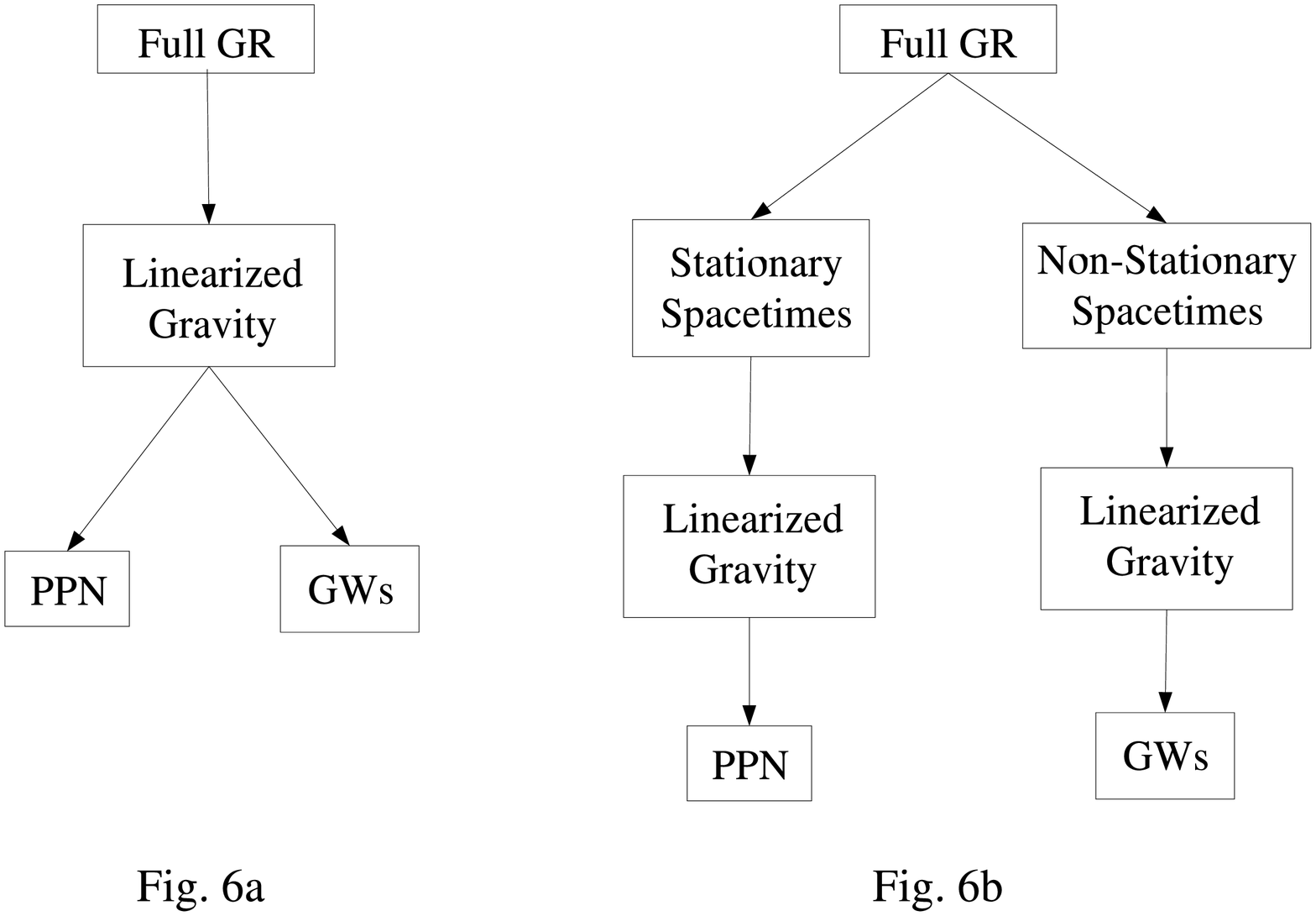}
\end{center}
\caption{\label{Fig-7}Fig. 6a is a schematic of the hierarchy of
approximations that we usually associate with GR. Fig. 6b is the
modified hierarchy based on the results of this paper. Notice the
explicit separation between stationary and nonstationary spacetimes
from the start.}
\end{figure}

Traditionally, there is an almost linear progression of
approximations, such as the hierarchy sketched in Fig.~\ref{Fig-7}a,
that we usually associated with GR. Within full GR we can
make the linearized gravity approximation, which is thought to be
inclusive of both the parametrized-post-Newtonian \cite{MTW}
(stationary) and the GW (nonstationary) limits (see also
\cite{Forward} and \cite{BCT}). We can, with further restrictions,
then pass over to either the parametrized-post-Newtonian (PPN) limit
or the GW limit within this linearized approximation. Based on the
arguments above and the results of our analysis in this paper, we find
the separation between the PPN and GW limits to occur at a much higher
level; the dynamics of particles in stationary spacetimes is
drastically different from that in nonstationary spacetimes. This
conclusion is consistent with \cite{hawking}, \cite{BD} and
\cite{Wald_0}, all of whom note differences between dynamical theories
in stationary versus nonstationary spacetimes. Thus, instead of the
standard formulation consisting of the linear sequence of
approximations in Fig.~\ref{Fig-7}a, we should, as shown in
Fig.~\ref{Fig-7}b, instead separate the dynamics in stationary versus
nonstationary spacetimes from the start.

What is most surprising about our results is the form that the
coupling of general GWs to the test particle takes in the GLF. The
tidal nature of the force that a GW induces on a test particle
introduces an effective velocity field $N_a$ to the system that
couples \textit{vectorially}, under the Lorentz Group, to
the particle, even though the GW itself is a \textit{rank-2}
tensor. This $N_a$\textemdash being a path integral\textemdash is in
general a \textit{nonlocal} function of the of Ricci
coefficients, although for planar GWs it reduces to the product of the
Ricci coefficient and the position of the test particle. [A vector
coupling of GWs to matter was proposed by one of the present authors
(R.Y.C.) in \cite{Chiao} based on a generalization of the PPN
formalism.] Thus the motion of the test 
particle is determined by the cumulative effects of the GW along the
distance \textit{between} the observer's worldline and the test
particle's. Like the vector potential for EM, $N_a$ is a
transverse field satisfying the wave equation, and like EM, effective
electric $\mathfrak{E}$ and magnetic $\mathfrak{B}$ fields can also be
constructed from $N_a$ that obey equations that have the same form as
Maxwell's equations. Thus, $N_a$ functions as a vector potential for
$\mathfrak{E}$ and $\mathfrak{B}$ with its gauge group being the local
Galilean group. Indeed, the equations of motion for the test particle
in the presence of arbitrary GWs has the form of the Lorentz force
with the mass $m$ of the particle playing the role of the ``charge.''

Surprising though this vector coupling may be at first glance,
after further study we see that there are deep
connections between $N_a$ and the tetrad formulation of eneral
relativity. $\mathfrak{E}$ and $\mathfrak{B}$ are proportional to the
path integrals of the electric $E_{ab}$ and magnetic $H_{ab}$ parts of
the Weyl tensor, respectively, and it is precisely the Weyl tensor that
contains GW excitations. It is also known that $E_{ab}$ and $H_{ab}$
obey tensor Maxwell's equations for GWs propagating on a flat,
source-free background, and the fact that $\mathfrak{E}$ and
$\mathfrak{B}$ satisfy Maxwell's equations is simply the reflection
that $E_{ab}$ and $H_{ab}$ do. In fact, the identity
$\epsilon^{ABCD}\partial_B\mathfrak{F}_{CD}=0$ holds precisely because
of the 1st and 2nd Bianchi conditions for the Riemann curvature tensor,
while $\partial^A\mathfrak{F}_{AB}=0$ only holds because we are in a
source-free region for GWs. As for $N_a$ itself, it is precisely
``half'' of the loop integral of the real part of Ashtekar's
connection in the loop-variable formulation of quantum
gravity. Indeed, Ashtekar's loop integrals can be thought of as
quantum corrections to the ``classical'' $N_a$ presented here
\cite{Ashtekar}.

Since $N_a$ is a nonlocal object, it is natural to question the
practicality of Eq.~$(\ref{60})$ as compared to, say,
Eq.~$(\ref{61})$, whose form for long-wavelength GWs is well
known. Indeed, it would seem that for Earth-based systems the
long-wavelength version of Eq.~$(\ref{61})$ would be sufficient to
describe all physically interesting systems. We note, however, that at
$4$ km the arms of LIGO are comparable to the reduced wavelength of
the GWs for GWs with a frequency of 10 khz, which is at the upper limits of
LIGO's frequency response spectrum \cite{LIGO-Report};
measurements of the spatial variation of GWs are becoming
experimentally accessible. Similarly, when LISA's (\textit{L}aser
\textit{I}nterferometer \textit{S}pace \textit{A}ntenna) arm lengths
begins to approach the reduced wavelengths of GWs in the
high-frequency limit of its sensitivity, the concept of the $N_a$
field and the GLF becomes essential. For these arbitrary GWs,
Eq.~$(\ref{60})$ instead of Eq.~$(\ref{61})$ must be used. Since it is
$N_a$ in general that will be measured, and not $h_{ab}$, it is an
interesting, and important, question as to how much information about
$h_{ab}$ can be obtained from measurements of $N_a$.

While most of our focus has been on classical dynamics,
we end this paper by looking at a gedanken experiment that probes the
essential difference between the classical and quantum dynamics of
nonrelativistic test particles in the linearized gravity limit.
Extension of Eq.~$(\ref{37})$ to quantum mechanics is
straightforward. For nonrelativistic, neutral particles with
wave function $\psi$,
\begin{equation}
i\hbar\frac{\partial\psi}{\partial t} =
-\frac{\hbar^2}{2m}\nabla^2\psi -i\hbar N_a\partial^a\psi.
\label{a}
\end{equation}
Notice that the interaction term $N_a\partial^a\psi$ has the same form
as the coupling of an EM vector potential $A_B$ (expressed in GLF
coordinates) to a charge particle coupled to weak EM fields. In the EM
case, the solution of equations of this form leads, on the quantum
level, to the Aharonov-Bohm effect, which is expressible through
Yang's nonintegrable phase factor (Wilson loop) \cite{Yang}
$\lambda_{EM}=\exp\{i(e/\hbar)\oint A_B dX^B\}$. In
direct analogy, Eq.~$(\ref{a})$ suggests that like the EM case,
a \textit{time-dependent} Aharanov-Bohm interference experiment such
as those described in \cite{Stod} or \cite{Anan} can in principle be
done, but now with GWs. We would expect this to lead to a
corresponding phase factor for gravity $\lambda_{GR} =
\exp\{i(m/\hbar) \oint N_A dX^A\}$. While actually performing this
experiment would be unrealistic at this time, the corresponding
gedanken experiment does illustrate both the essential difference
between the classical and quantum systems, and the importance of this
gravitational nonintegrable phase factor.

\begin{figure}
\begin{center}
\includegraphics[width=1\textwidth]{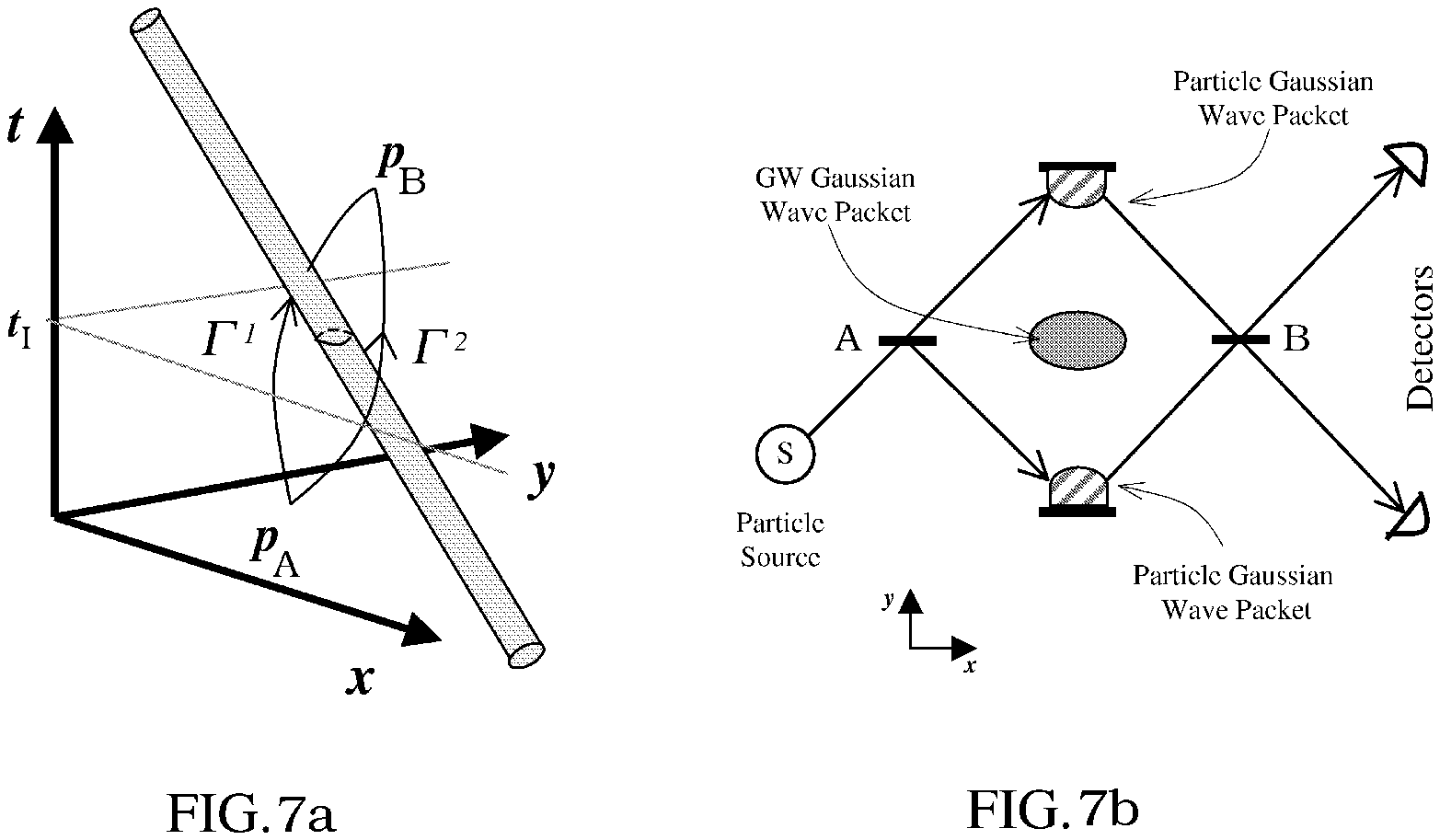}
\end{center}
\caption{\label{Fig_8} Fig.~7a is a sketch of spacetime diagram
showing the passage of the unipolar, Gaussian GW (shown in gray) and
the worldlines of two particles beams, along the two paths $\mathit{\Gamma}^1$
and $\mathit{\Gamma}^2$ (i.~e.~the two dominant Feynman paths), of the
interferometer encircling it. Events $p_A$ and $p_B$ correspond to the
splitting and recombining of the initial and final beam splitters
shown in Fig.~7a. Fig.~7b shows a schematic of a possible experimental
set-up (a Bonse-Hart interferometer) at the given time slice $t_I$
shown in Fig.~7a. The GW beam passes through the center of the
Bonse-Hart interferometer, and the gray oval shows the cross section
of the Gaussian beam waist which forms at the center of the plane of
the interferometer.}
\end{figure}

Consider the gedanken Aharanov-Bohm-type experiment shown in
Fig.~$\ref{Fig_8}$ using a Gaussian GW packet with a width $l_{GW}$ at
some time $t_I$. Fig.~$\ref{Fig_8}$a shows the spacetime diagram for the
particle and the GW used in the interferometry experiment with a
time slice drawn in at $t_I$. A schematic of the corresponding
physical apparatus at this time slice is shown in
Fig.$\ref{Fig_8}$b. We require that the GW also be an unipolar
wave packet so that at all times the amplitude of the $\mathfrak E$
and $\mathfrak B$ for the packet is non-negative. The interfering
particle is taken to be a Gaussian wave packet as well, but with width
$l_p$ at any time $t$. After being emitted at the source, the
particle at event $p_A$ passes through the first beam splitter shown
in Fig.~$\ref{Fig_8}$b, and there is a finite probability amplitude it
will propagate \textit{either} along $\mathit{\Gamma}^1$ \textit{or} along
$\mathit{\Gamma}^2$ until it is is recombined at event $p_B$. It
is impossible, even in principle, to know \textit{which path} the
particle took through the interferometer. As usual, the combined path
$\mathit{\Gamma} = \mathit{\Gamma}^1 \cup \mathit{\Gamma}^2$ encircles
the GW beam.

The Bonse-Hart interferometer of a type shown in Fig.~$\ref{Fig_8}$b is a
concrete example of an interferometer that could be used in this
experiment; in this case the interfering particle could be
neutrons. The gray oval patch at the center of the interferometer
represents the time slice at $t_I$ shown in Fig.~$\ref{Fig_8}$a of the
Gaussian beam waist. Most
importantly, we choose the size of the interferometer such that
$R>>l_{GW} + l_p$ where $R$ is the distance from the center of the GW
beam in Fig.~$\ref{Fig_8}$b to each arm of the interferometer. Thus,
on a classical level the GW and the neutron worldlines do not
intersect, and therefore the neutron feels no forces at
any time arising from the GW. On a quantum mechanical level, the
amplitude of the neutron's wave function is exponentially small where
the amplitude of the GW is large, and the amplitude of the GW is
exponentially small where the amplitude of the wave function is large.

Although the wave function of the  neutrons satisfy Eq.~$(\ref{a})$,
because they are Gaussian wave packets we can make a WKB-like
approximation by taking $\psi \approx e^{i\Theta}$. For $h_{ab}$  with
small spatial variations along the worldlines, Eq.~$(\ref{a})$ reduces to
\begin{equation}
0 = \frac{\partial\Theta}{\partial t} + N_a\partial^a\Theta,
\label{b}
\end{equation}
where we have neglected terms $\mathcal{O}(\vert\nabla\Theta\vert^2)$
since $N_a$ is small. The solution of Eq.~$(\ref{b})$ in the linearized
gravity limit for the particle propagating along worldline
$\mathit{\Gamma}^1$ is
\begin{equation}
\Theta_{\mathit{\Gamma}^1} = \frac{m}{\hbar} \int_{\mathit{\Gamma}^1} N_A dX^A,
\label{c}
\end{equation}
from Eq.~$(\ref{75})$ of Appendix B. Similarly, if the particle had
traveled the worldline $\mathit{\Gamma}^2$,
\begin{equation}
\Theta_{\mathit{\Gamma}^2} = \frac{m}{\hbar} \int_{\mathit{\Gamma}^2} N_A dX^A.
\label{d}
\end{equation}
The phase factor
\begin{equation}
\exp\{i\Delta\Theta\} \equiv
\exp\left\{\Theta_{\mathit{\Gamma}^1}-\Theta_{\mathit{\Gamma}^2}\right\} =
\exp\left\{i\frac{m}{\hbar}\int_{\mathit{\Gamma}} N_A dX^A\right\},
\label{e}
\end{equation}
is a measure of the phase difference between the particle propagating
along $\mathit{\Gamma}^1$ versus propagating along
$\mathit{\Gamma}^2$. Equation $(\ref{e})$ is simply the
$\lambda_{GR}$ that we predicted above. Next, since $\mathit\Gamma$ is
a closed loop, by the Stokes's theorem,
\begin{equation}
\exp\{i\Delta\Theta\} =
\exp\left\{i\frac{m}{\hbar}\int_{\mathcal
  D}\mathfrak{F}_{AB}dX^A\wedge dX^B\right\},
\label{f}
\end{equation}
where $\mathcal D$ is the surface bound by $\mathit\Gamma$ and we have
used the definition of $\mathfrak{F}_{AB}$. Because the boundary of
$\mathcal D$ is made up of timelike curves, from Eqs.~$(\ref{47})$,
$(\ref{54})$, and $(\ref{56})$ the dominant contribution to the surface
integral in Eq.~$(\ref{f})$ comes from $R_{0a,0b}$. As expected, the
phase factor depends on the Riemann curvature tensor, and is
independent of coordinate choice, even though Eq.~$(\ref{a})$ is
dependent on the gauge-dependent field $N_a$.

We note also the underlying topological nature of
Fig.~$\ref{Fig_8}a$. If a path $\mathit\Gamma$ is chosen that does not
encircle the GW beam, $\Delta\Theta=0$; if it does, $\Delta\Theta\ne
0$. $\mathit\Gamma$ cannot be shrunk to zero without cutting though the GW
beam, and thus altering the topology of the $\mathit\Gamma$-GW beam
system. Moreover, each time one goes around $\mathit\Gamma$, the phase
difference $\Delta\Theta$ changes by an integral multiple of the same
factor; $\Delta\Theta$ is thus proportional to the linking number of
$\mathit\Gamma$ around the GW beam. That $\Delta\Theta$ is related to the
linking number is not unexpected since $N_a$ is related to Ashtekar's
loop variables. Nonetheless, it does point out the possibility of
doing experiments such as \cite{Tono} that directly measures the
linking number of $\mathit\Gamma$, and this indicates the underlying
topological nature of this gravitational Aharonov-Bohm-type effect.

\appendix{}

\section{Review of Linearized Gravity and Tetrad Formalisms}

In this appendix, we present a brief review of some of the properties of
general tetrad frames, differential forms, and linearized gravity that
we shall need. The review is not exhaustive and the reader is referred
to \cite{Wald} or \cite{deF} for a more complete presentation.

A tetrad is a local coordinate system formed by a set of four
orthonormal vectors $\{e_A^{\>\>\>\>\mu}\}$ such that $e_A^{\>\>\>\>\mu}
e_{B\mu} = \eta_{AB}$ and $e^A_{\>\>\>\>\mu}e_{A\nu} =
g_{\mu\nu}$. These $\{e_A^{\>\>\>\>\mu}\}$ do not have to
be tied to any observer's worldline as we have done in the above, and
the results presented here are valid in general.

Since $e_A^{\>\>\>\>\nu}$ is a unit vector, the covariant derivative
$\nabla_\mu e_{B\nu}$ can only be a rotation or boost of
$e_B^{\>\>\>\>\nu}$. This ``rotation matrix'' is determined by the
Ricci rotation coefficients:
\begin{equation}
\omega_{CAB} = e_C^{\>\>\>\>\mu} e_A^{\>\>\>\>\nu}\nabla_\mu e_{B\nu},
\label{62}
\end{equation}
with $\omega_{CAB} = - \omega_{CBA}$. It is straightforward to
show that in a tetrad frame,
\begin{eqnarray}
R_{ABCD} =&{}& e_A^{\>\>\>\>\mu}\partial_\mu\omega_{BCD} -
e_B^{\>\>\>\>\mu}\partial_\mu\omega_{ACD} -
\nonumber \\
&{}&\left\{\omega_{AFC}\omega_B{}^F_{\>\>\>\>D} -
\omega_{BFC}\omega_A{}^F_{\>\>\>\>D} +
\omega_{AFB}\omega^F_{\>\>\>\>CD} - \omega_{BFA}\omega^F_{\>\>\>\>CD}
\right\}.
\label{63}
\end{eqnarray}
We emphasize that objects with capital Roman indices in the
tetrad frame\textemdash such as $\omega_{CAB}$\textemdash are
\textit{scalars}. They only take the partial derivative, and not the
covariant derivative.

Given a tetrad frame, it is natural to work with differential
forms, which we shall denote by symbols in boldface.
As $1$-forms we have $\bm{e}_A \equiv e_{A\mu}\bm{d}x^\mu$
and $\bm{\omega}_{AB} = \omega_{\mu AB} \bm{d}x^\mu$, and as 2-forms
we have $\bm{R}_{AB} \equiv R_{\mu\nu AB} \bm{d}x^\mu \wedge
\bm{d}x^\nu/2$. As usual, $\bm{d}$ is the exterior
derivative. Equation $(\ref{62})$ then becomes
\begin{equation}
\bm{de}_A = \bm{e}^B\wedge \bm{\omega}_{AB},
\label{64}
\end{equation}
where $\wedge$ is the wedge product; $\bm{\omega}_{AB}$'s role as a
``rotation'' or boost matrix is now manifest. Equation $(\ref{63})$ is then
\begin{equation}
\bm{R}_A^{\>\>\>\>B} = \bm{d\omega}_A^{\>\>\>\>B} +
\bm{\omega}_A^{\>\>\>\>C}\wedge \bm{\omega}_C^{\>\>\>\>B}.
\label{65}
\end{equation}
Taking the exterior derivative of Eq.~$(\ref{64})$, we get the 1st
Bianchi identity,
\begin{equation}
0= \bm{e}_B\wedge \bm{R}_A^{\>\>\>\>B},
\label{66}
\end{equation}
and the exterior derivative of Eq.~$(\ref{66})$ gives the 2nd Bianchi
identity
\begin{equation}
\bm{dR}_A^{\>\>\>\>B} = \bm{R}_A^{\>\>\>\>C}\wedge
\bm{\omega}_C^{\>\>\>\>B}
-\bm{\omega}_A^{\>\>\>\>C}\bm{R}_C^{\>\>\>\>B}.
\label{67}
\end{equation}

In the case of linearized gravity, $g_{\mu\nu} = \eta_{\mu\nu} +
h_{\mu\nu}$ where $h_{\mu\nu}$ is ``small'', and we are only
concerned with terms linear in $h_{\mu\nu}$. Then $g^{\mu\nu} =
\eta^{\mu\nu} - h^{\mu\nu}$ where on the left hand side we use
$\eta_{\mu\nu}$ to raise and lower indices. In this limit
\begin{eqnarray}
\Gamma^\alpha_{\mu\nu} &=&
            \frac{1}{2}\left(\partial_\mu h_\nu^\alpha
+ \partial_\nu h^\alpha_\mu -\partial^\alpha h_{\mu\nu}\right),
\nonumber \\
R^\mu_{\>\>\>\>\nu\alpha\beta}& =& \frac{1}{2}
                \left\{
                \partial_\nu(
                    \partial_\alpha h_\beta^\mu
                    -
                    \partial_\beta h_\alpha^\mu)
                +
                \partial^\mu
                (\partial_\beta h_{\nu\alpha}
                -
                \partial_\alpha h_{\nu\beta})
                \right\},
\nonumber \\
R_{\mu\nu} &=& -\frac{1}{2}\Box h_{\mu\nu} + \frac{1}{2}
        \partial_\mu\partial^\alpha h_{\alpha\nu} +\frac{1}{2}
        \partial_\nu\partial^\alpha h_{\alpha\mu} - \frac{1}{2}
        \partial_\mu\partial_\nu h_\alpha^\alpha,
\nonumber \\
R &=& -\Box h_\mu^\mu + \partial_\mu \partial_\nu h^{\mu\nu}.
\label{68}
\end{eqnarray}
For GWs in the TT gauge,
$h_\mu^\mu=0$, $\partial^\mu h_{\mu\nu} =0$, and $h_{0\mu}=0$, so that
$R=0$, $R_{0\mu}=0$, and the equation of motion for GWs is
$R_{\hat{a}\hat{b}} = -\Box h_{\hat{a}\hat{b}}/2=0$.

To construct the tetrad frame for linearized gravity, we first note
that in flat spacetime the tetrad frame is trivial:
$\{\delta_A^\mu\}$. The presence of small $h_{\mu\nu}$
rotates these vectors and
\begin{equation}
e_A^{\>\>\>\>\mu} = \left(\delta^\mu_\nu - \frac{1}{2}h^\mu_\nu\right)
\delta_A^\nu,\qquad
e^A_{\>\>\>\>\mu} = \left(\delta_\mu^\nu + \frac{1}{2}h_\mu^\nu\right)
\delta^A_\nu.
\label{69}
\end{equation}
Note that this choice is \textit{not} unique. As noted in Sec.~IV
4, given any tetrad frame, we can always do a \textit{local} Lorentz
transformation that will still preserve the orthonormality of
$\{e_A^{\>\>\>\>\mu}\}$. The choice we have made for
$\{e_A^{\>\>\>\>\mu}\}$ for the GLF, and used in the
construction in Sec.~I, corresponds to an observer at
rest in his proper frame.

Using Eq.~$(\ref{62})$ and the Levi-Civita connection in
Eq.~$(\ref{68})$, we find that
\begin{equation}
\omega_{CAB} = -\frac{1}{2} \left(\partial_A h_{BC}
- \partial_B h_{AC}\right),
\label{70}
\end{equation}
while $R_{ABCD} =\partial_A\omega_{BCD} - \partial_B\omega_{ACD}$ in
the linearized gravity limit. Also in this limit, the equations for
the Ricci tensor and scalar in Eq.~$(\ref{68})$ have the same form in
the tetrad frame, with the replacement of Greek indices by capital
Roman indices.

\section{Phase Factor Solution}

Equation $(\ref{b})$ is a quasi-linear partial differential equation
\cite{handbook} whose method of solution is well known. First, both
$t$ and $X^a$ are considered functions of a parameter $\tau$, so that
$\Theta = \Theta(t(\tau), X^a(\tau))$, and defines a constant surface
in $(t,X^a)$ space. Consequently,
\begin{equation}
0 = \frac{dt}{d\tau} \frac{\partial\Theta}{\partial t}
+ \frac{dX^a}{d\tau} \partial_a\Theta,
\label{71}
\end{equation}
and for Eq.~$(\ref{71})$ to be a solution of Eq.~$(\ref{b})$,
\begin{equation}
\frac{\partial t}{\partial\tau} = 1, \qquad \frac{d X^a}{d\tau} = N^a;
\label{72}
\end{equation}
the quasi-linear partial differential equation reduces to
the solution of a set of ordinary differential equations. Although
Eq.~$(\ref{72})$ can be solved using standard methods once an initial
condition is given, the underlying physics become much clearer if we consider
instead the following function
\begin{equation}
\Theta_{\widetilde{\mathit\Gamma}}(t(T), X^a(T)) = \frac{m}{\hbar}
\int_{\widetilde{\mathit\Gamma}}
N_A(t(\tau), X^a(\tau))
dX^A,
\label{73}
\end{equation}
where the integral is from a fixed point $(t(0), X^a(0))$ to the point
$(t(T), X^a(T))$ along $\widetilde{\mathit\Gamma}$. The prefactor
$m/\hbar$ is included so that $\Theta_{\widetilde{\mathit{\Gamma}}}$ is
unitless. We choose $\widetilde{\mathit\Gamma}$ such that its tangent
vector is given by Eq.~$(\ref{72})$, and we parameterize it by $\tau\in
[0,T]$. $\widetilde{\mathit\Gamma}$ starts at the point $(t(0),
X^a(0))$, and is used also as the initial condition for
Eq.~$(\ref{72})$. Clearly,
\begin{equation}
\frac{d\Theta_{\widetilde{\mathit\Gamma}}(t, X^a)}{d\tau} =\frac{m}{\hbar}
N_A\frac{dX^A}{d\tau} = \frac{m}{\hbar} N_aN^a \approx 0,
\label{74}
\end{equation}
in the linearized gravity limit, while the left-hand-side vanishes
identically from Eq.~$(\ref{71})$. Thus,
$\Theta_{\widetilde{\mathit\Gamma}}(t,X^a)$ is a solution of
Eq.~$(\ref{b})$. Because it is path dependent, this solution is not
unique, however.

The spatial component of the tangent vector to
$\widetilde{\mathit\Gamma}$ lies along $N_a$, and for plane waves,
$N_a$ is perpendicular to the direction of propagation. Thus, for the
Gaussian beam in Fig.~$\ref{Fig_8}a$, $\widetilde{\mathit{\Gamma}}$
will wrap around the beam. We can,
of course, go around this beam either in a clockwise or
counterclockwise direction, and we denote a clockwise path by
$\widetilde{\mathit\Gamma}^{-}$ and a counterclockwise path by
$\widetilde{\mathit\Gamma}^{+}$. To $\widetilde{\mathit\Gamma}^{-}$
there is then the corresponding the function
$\Theta_{\widetilde{\mathit\Gamma}^{-}}$, and to
$\widetilde{\mathit\Gamma}^{-}$ there is the corresponding function
$\Theta_{\widetilde{\mathit\Gamma}^{+}}$.

Consider now the functions
\begin{eqnarray}
\Theta_{\mathit{\Gamma}^1}(t(T), X^a(T))&=& \frac{m}{\hbar}
\int_{\mathit{\Gamma}^1} N_A dX^A, \nonumber \\
\Theta_{\mathit{\Gamma}^2}(t(T),X^a(T))&=& \frac{m}{\hbar}
\int_{\mathit{\Gamma}^2} N_A dX^A,
\label{75}
\end{eqnarray}
and we restrict ourselves to those $\widetilde{\mathit\Gamma}^{\pm}$
that lie on the surface of the beam, meaning that
$\widetilde{\mathit\Gamma}^{\pm}$ do not come closer than $l_{GW}$ to
the center of the beam of GWs. Because $N_a$ is exponentially small
outside of the Gaussian beam, $\Theta_{\mathit{\Gamma}^1} \approx
\Theta_{\widetilde{\mathit\Gamma}^{+}}$ and
$\Theta_{\mathit{\Gamma}^2} \approx
\Theta_{\widetilde{\mathit\Gamma}^{-}}$. Thus, Eqs.~$(\ref{75})$ are
solutions of Eq.~$(\ref{b})$.

\nonumber

\begin{acknowledgments}

A.D.S.~and R.Y.C.~were supported by a grant from the Office of Naval
Research.  We thank John Garrison and Jon Magne Leinaas for many
clarifying and insightful discussions. We also thank William Unruh for
reading a very early draft of this paper, and Robert Herrlich for
editing the final draft.

\end{acknowledgments}


\end{document}